\documentclass[12pt,preprint]{aastex}
\newcommand{\vnn}{\mbox{N}}
\begin{document}
\title{SZE Signals in Cluster Models}
\author{Beth A. Reid}
\affil{Department of Physics}
\affil{Princeton University, Princeton, NJ 08544}
\email{breid@princeton.edu}
\and
\author{David N. Spergel}
\affil{Department of Astrophysical Sciences}
\affil{Princeton University, Princeton, NJ 08544}
\email{dns@astro.princeton.edu}
\begin{abstract}
%
%
The upcoming generation of SZE surveys will shed fresh light onto the study of clusters.  What will this new observational window reveal about cluster properties?  What can we learn from combining X-ray, SZE, and optical observations?  How do variations in the gas entropy profile, dark matter concentration, accretion pressure, and intracluster medium (ICM) mass fraction affect SZE observables?  We investigate the signature of these important cluster parameters with an analytic model of the ICM.  Given the current uncertainties in ICM physics, our approach is to span the range of plausible models motivated by observations and a small set of assumptions.  We find a tight relation between the central Compton parameter and the X-ray luminosity outside the cluster core, suggesting that these observables carry the same information about the ICM.  The total SZE luminosity is proportional to the thermal energy of the gas, and is a surprisingly robust indicator of cluster mass: $L_{\rm SZ} \propto f_{\rm ICM} M^{5/3}$.  We show that a combination of $L_{\rm SZ}$ and the half-luminosity radius $r_{\rm SZ}$ provides a measure of the potential energy of the cluster gas, and thus we can deduce the total energy content of the ICM.  We caution that any systematic variation of the ICM mass fraction will distort the expected $L_{\rm SZ} - M$ calibration to be used to study the evolution of cluster number density, and propose a technique using kSZ to constrain $f_{\rm ICM}(M,z)$.\\
\end{abstract}
%
\section{\bf Introduction}
\label{intro}
As the largest and most recently formed relaxed objects in the universe, clusters provide cosmological constraints complementary to other observations.  \citet{press/schechter:1974} showed that their co-moving number density is exponentially sensitive to both cluster mass and the amplitude of the linear power spectrum of density fluctuations.  The number density of massive galaxy clusters in the local universe constrains a model-dependent combination of the amplitude of fluctuations at $8 \; h^{-1}$ Mpc ($\sigma_{8}$) and the matter density ($\Omega_{\rm m}$).  The redshift evolution of cluster number counts depends on the linear growth factor and and the co-moving volume element, and can thereby potentially break the degeneracy in the family of $\Lambda$ and quintessence models allowed by primary CMB anisotropy \citep{wang/steinhardt:1998}.  X-ray and optical cluster counts at low and intermediate redshifts have been used to constrain cosmological parameters \citep[for recent results, see][]{allen/etal:2003, allen/etal:2004a, bahcall/etal:2003, henry:2004, ikebe/etal:2002, pierpaoli/etal:2003, rapetti/allen/weller:2005, reiprich/bohringer:2002, rosati/borgani/norman:2002, shimizu/etal:2003, viana/nichol/liddle:2002, vikhlinin/etal:2003}.\\
The Sunyaev-Zel'dovich effect (SZE) flux is an excellent probe of high redshift clusters, and many SZE surveys are in development, such as ACT\footnote{http://www.hep.upenn.edu/act/}, APEX\footnote{http://bolo.berkeley.edu/apexsz/index.html}, Planck\footnote{http://www.rssd.esa.int/index.php?project=PLANCK\&page=index}, and SPT\footnote{http://spt.uchicago.edu/}.  While the information from high redshift available in SZE surveys can potentially provide tight cosmological constraints \citep{haiman/mohr/holder:2001}, recent work suggests that uncertainties in the mass-observable relation and its scatter can severely degrade their sensitivity.  Fortunately, internal calibration and some follow-up observations can recover much of a survey's sensitivity {\em if} the mass-observable relation and its scatter can be described accurately in the relevant redshift range by a reasonably small set of parameters (\citealt{hu:2003, majumdar/mohr:2004, lima/hu:2004, lima/hu:2005}; but see also \citealt{francis/bean/kosowsky:prep}).  The results presented in this paper support their assumption: we find a tight mass-observable SZE relation that is robust to a wide range of model variations.\\
The SZE is a distinct spectral signature in the CMB due to Thomson scattering of CMB photons and hot electrons with a magnitude proportional to the integrated electron pressure along the line of sight \citep[for discussions of SZE physics, see][]{birkinshaw:1999, carlstrom/holder/reese:2002}.  A unitless measure of the effect is the Compton parameter $y$ \citep{sunyaev/zeldovich:1972}:
\begin{equation}
\label{compton}
y = \frac{\sigma_{Thomson}}{m_e c^{2}} \int P_{e} \; dl.
\end{equation}
We define the SZE luminosity $L_{\rm SZ}$ as the integration of $y$ over the projected cluster area, so that 
\begin{equation}
\label{SZflux}
f_{\rm SZ} d_{A}^{2} = L_{\rm SZ} = \int y \; dA = \frac{\sigma_{Thomson}}{m_e c^{2}} \int P_{e} dV = \frac{\sigma_{Thomson}}{m_e c^{2}} N_{e} \langle T_{e} \rangle.
\end{equation}
Assuming local thermal equilibrium, $L_{\rm SZ}$ directly measures the total thermal energy content of the intracluster medium (ICM), and $y(\theta)$ informs us on how the gas is distributed in the cluster.  Because $d_{A}$ flattens at intermediate redshift, the SZE flux $f_{\rm SZ}$ becomes approximately redshift independent.\\
The self-similar collapse model for the ICM considers only gravitational physics and predicts scaling relations between cluster properties.  X-ray observations indicate significant deviations from these predictions, most notably in the X-ray luminosity-temperature relation \citep{markevitch:1998, arnaud/evrard:1999}.  Many complex non-gravitational processes determine the final thermal state of the ICM.  We discuss many of them in Sec.~\ref{xraysec} along with related X-ray observations.   In Sec.~\ref{results} we introduce an analytic model of smooth accretion \citep{voit/etal:2003} that provides a conceptual foundation for determining the gas properties.  We then relax some of the assumptions of \citet{voit/etal:2003} and parametrize our uncertainties in ICM physics: the poorly understood effects of heating, cooling, and conduction is encoded in the gas entropy profile and the ICM mass fraction, variation in the dark matter potential is characterized by the concentration parameter, and the boundary condition is set by an accretion pressure.  Motivated by the physics and observations discussed in Sec.~\ref{xraysec}, our main goal is to explore SZE and X-ray observables over a plausible range of parameter values using a set of phenomenological models under the assumptions of spherical symmetry and hydrostatic equilibrium.   We compare the resulting profiles to recent observations of nearby, relaxed clusters reported in \citet{vikhlinin/etal:2005}.  We quantify deviations from the expected $L_{\rm SZ} - M$ relationship and indicate the main sources of scatter accessible to our models.  In Sec.~\ref{relations} we describe relations between cluster properties and observables that hold across the range of models we consider.  In particular, we show that SZE observables alone provide a robust measure of the total energy content of the ICM.  We discuss the implications and limitations of our models in Sec.~\ref{conc}, and focus on the important point of constraining the ICM cluster mass fraction $f_{\rm ICM}$ using the kinetic Sunyaev-Zel'dovich effect (kSZ) in Sec.~\ref{fbdiscussion}.  Sec.~\ref{future} states our conclusions and the implications of this work for future SZE surveys.\\
%
\section{\bf ICM Physics}
\label{xraysec}
In the self-similar spherical collapse model \citep{peebles:POPC} and in agreement with N-body simulations, all virialized clusters at fixed redshift have the same overdensity relative to the background.  In the absence of non-gravitational processes, we expect the gas temperature to be set by the dark matter virial temperature: $T_{\rm vir} = \frac{GM\mu m_{p}}{2R} \propto M^{2/3}$ where $\mu m_{p}$ is the mean gas particle mass.  These assumptions also predict $L_{X} \propto T^{2}$ when thermal bremsstrahlung dominates X-ray emission ($T \gtrsim 2 \; {\rm keV}$) and $L_{X} \propto T$ at lower temperatures when line emission dominates.  However, observations indicate $L_{X} \propto T^{2.6-2.9}$ \citep{markevitch:1998, arnaud/evrard:1999}.  This deviation from the expected scaling has instigated extensive theoretical investigation of the effects of various non-gravitational processes.  Preheating of the intergalactic medium at an early time \citep{kaiser:1991}, radiative cooling \citep{voit/bryan:2001}, cooling with energy injection due to feedback \citep{ostriker/bode/babul:2005}, and quasar activity \citep{lapi/cavaliere/menci:2005} can all reproduce the observed $L_X-T$ relation.  Fortunately, the latest high resolution X-ray observations provide new constraints for ICM models out to nearly half the dark matter virial radius \citep{vikhlinin/etal:2005}.\\
Under the assumptions of spherical symmetry and hydrostatic equilibrium, X-ray surface brightness and temperature profiles allow extraction of a cluster's total mass profile.  Recent data from nearby, relaxed clusters \citep{vikhlinin/etal:2005,pratt/arnaud:2005} are consistent with the universal NFW profile \citep{navarro/frenk/white:1997} with $\Lambda CDM$ best fit concentration parameters and scatter \citep{dolag/etal:2004}.  In Vikhlinin et al.'s sample of 13 clusters with temperature range 0.7 - 9 keV, temperature profiles are self-similar only at $r \gtrsim 0.15 r_{\rm vir}$.  However, using the gas mass weighted rather than X-ray emission weighted temperature, they still find agreement with self-similar predictions for the $M-T$ scaling.\\
Since cluster gas evolves nearly adiabatically, the entropy profile is a useful description of the gas: once the entropy distribution is fixed, convective stability, a confining pressure at the cluster gas boundary, and the assumption of hydrostatic equilibrium determine the gas profile in a dark matter potential.  The customary definition of `entropy' in this field is \citep{tozzi/norman:2001,voit/etal:2003}
\begin{equation}
\label{entropyeqn}
K = \frac{P}{\rho^{5/3}} = \frac{T}{\mu m_p \rho^{2/3}} \propto e^{2s/3}
\end{equation}
and is related to the thermodynamic entropy per particle $s$ as given above.  Typically, one considers the function $K(M_{\rm gas})$, where $M_{\rm gas}$ is the mass of gas with entropy is $\leq K(M_{\rm gas})$.  A useful quantity is the characteristic entropy of a region of overdensity $\Delta$ as defined by
\begin{equation}
\label{charentropy}
K_{\Delta} = \frac{T_{\Delta}}{\mu m_{p}} \left(\Delta f_{\rm b} \rho_{cr}\right)^{-2/3}
\end{equation}
with $T_{\Delta} =  \frac{GM_{\Delta} \mu m_{p}}{2r_{\Delta}}$ the characteristic temperature of the dark matter potential at overdensity $\Delta$, and $f_{\rm b}$ the universal baryon fraction.  The self-similar collapse model predicts a universal entropy profile $K(r) \propto r^{1.1}$ fixed by the physics of gravitational collapse \citep{tozzi/norman:2001,borgani/etal:2001b}, with the normalization scaling as $K_{\Delta} \propto T_{\Delta}$ for clusters at fixed redshift.  Observed entropy profiles \citep{pratt/arnaud:2005} are approximately self-similar down to 2 keV in the radial range $0.05 r_{200} < r < 0.5 r_{200}$ with $K(r) \propto r^{0.94 \pm 0.14}$ \citep{pratt/arnaud:2005}, and they generally show a flatter core inside $0.1r_{200}$ \citep{pratt/arnaud:2003,ponman/sanderson/finoguenov:2003}.  The entropy profile normalization clearly deviates from the self-similar prediction and scales with cluster temperature as $K(0.1 r_{200}) \propto T^{0.65 \pm 0.05}$ \citep{ponman/sanderson/finoguenov:2003}.\\
As suggested by \citet{neumann/arnaud:2001}, similarity breaking in both the $L_{X} - T$ and $K - T$ relations are consistent with an ICM mass fraction temperature dependence, so that $f_{\rm ICM} \sim T^{1/2}$.  X-ray observations find evidence for lower gas densities in cooler clusters out to $\approx 0.35 r_{200}$ (\citealt{neumann/arnaud:2001, pratt/arnaud:2003, vikhlinin/etal:2005}; but see also \citealt{mathews/etal:2005}).  Current X-ray observations cannot measure gas profiles much past $0.5 r_{\rm vir}$, so the `missing' gas may be bound in the outer regions of the cluster, escaped from the cluster potential altogether, or condensed into intracluster stars \citep{gonzalez/zabludoff/zaritsky:2005, lin/mohr:2004}.  SZE signatures in WMAP also find that $f_{\rm ICM}$ increases with temperature \citep{afshordi/lin/sanderson:2005}.  Because $L_{\rm SZ} \propto f_{\rm ICM}$, any trend in the total ICM mass fraction will affect the $L_{\rm SZ} - M$ relation.  Such trends are quite possible, given trends with mass in observations of the cluster mass/near-infrared K-band luminosity relation \citep{lin/mohr/stanford:2004}, estimates of intracluster light \citep{lin/mohr:2004}, and in hydrodynamic simulations finding a mass-dependent ICM mass fraction \citep{kravtsov/nagai/vikhlinin:2005}.\\
\citet{voit/bryan:2001} have suggested that radiative cooling sets the entropy scale responsible for similarity breaking: gas below the cooling threshold is either condensed or injected with energy through feedback until it exceeds the cooling threshold.  Consistent with observed scaling relations, their simple model predicts for the central entropy $K \propto T^{2/3}$ for $T > 2 \; {\rm keV}$ and $K$ independent of $T$ for $T < 2 \; {\rm keV}$.  \citet{dave/katz/weinberg:2002} demonstrate a similar $K - T$ scaling in numerical simulations of groups that include radiative cooling and star formation.  In the analytic model of \citet{mccarthy/etal:2004} cooling approximately maintains the initial entropy power law, in agreement with the observed entropy gradients in clusters.\\
Cooling and subsequent feedback play an essential but enigmatic role in the thermodynamics of the ICM.  As many as $70 - 90\%$ of nearby clusters have cold cores of gas with $t_{cool}$ much smaller than the age of the cluster \citep{peres/etal:1998}, and similar results have been found in the redshift range $z \approx 0.15 - 0.4$ \citep{bauer/etal:2005}.  However, in several well-studied cold core clusters, \citet{tamura/etal:2001}, \citet{peterson/etal:2001}, \citet{sakelliou/etal:2002}, \citet{peterson/etal:2003}, and \citet{kaastra/etal:2004} find very little cooled gas below a quarter of the hot gas temperature.  Measurements of O VI emission in cold cores imply gas condensation rates well below the expectations from a simple cooling flow model \citep{bregman/etal:prep}, and altogether observations suggest subsonic energy injection spatially distributed in regions of $\sim 100$ kpc and continuous on timescales of $10^{8}$ years.  Thermal conduction and AGN activity may be important to balancing radiative cooling in cold cores \citep[see][for a comprehensive review of the cooling flow problem]{peterson/fabian:prep}.\\
A distinct source of feedback is needed to balance the severe overcooling present in cooling-only analytic calculations and cosmological hydrodynamic simulations \citep{oh/benson:2003, balogh/etal:2001}.  The stellar baryon fraction in clusters has been estimated at $5 - 20\%$  \citep{roussel/sadat/blanchard:2000, balogh/etal:2001, lin/mohr/stanford:2003}, though inclusion of intracluster stars may increase this fraction by a factor of two \citep{lin/mohr:2004, gonzalez/zabludoff/zaritsky:2005, zibetti/etal:2005}.  \citet{voit/bryan:2005} conclude that heating is needed at high redshift to solve this `cooling crisis,' and its resolution is most likely related to the high entropy levels observed in less massive clusters and groups.\\
%
%
A simple preheating of the intergalactic medium to a `universal entropy floor' by supernovae and AGN before the epoch of cluster formation \citep{kaiser:1991} introduces deviations from the expected scaling relations and mitigates the cooling crisis.  Heating occurring before accretion onto the cluster most efficiently raises the gas entropy level.  \citet{lapi/cavaliere/menci:2005} estimate $1/4 \; {\rm keV/particle}$ is available from supernovae, and is not enough to match the observed $L_{X}-T$ and $K-T$ relations.  \citet{lapi/cavaliere/menci:2005} also consider an additional $1/2 \; {\rm keV/particle}$ available for preheating from smooth, long-lived AGN outputs; this energy is sufficient to provide marginal agreement with the observed $L_{X}-T$ and $K-T$ scaling relations.  However, external preheating models with smooth accretion are generically disfavored observationally because they predict large isentropic cores in low mass systems \citep{tozzi/scharf/norman:2000,pratt/arnaud:2003,ponman/sanderson/finoguenov:2003} and require a large entropy floor ($\gtrsim 3 \times 10^{33} {\rm \; erg \; cm^2 \; g^{-5/3}}$) to match the $L_{X}-T$ relation \citep{mccarthy/etal:2004}.  We do consider this model in our computations for completeness.\\
Simulations show that the accretion process is far from spherical, often taking place along filaments.  Thus many researchers \citep{voit/etal:2003, ponman/sanderson/finoguenov:2003,voit/bryan:2005} have suggested a more favorable scenario of preheating where the effect of preheating is primarily to reduce the density of pre-shock gas accreting onto lower mass systems.  This raises the post-shock entropy normalization while producing the radial entropy profile expected from gravitational shock heating.  \citet{borgani/etal:2005} have explored this proposal in numerical simulations, and find that feedback must be better distributed than their strongest galactic winds in order to produce the desired levels of entropy amplification and radial gradient.  \citet{lapi/cavaliere/menci:2005} circumvent isentropic cores and introduce similarity-breaking through quasar blast waves within the cluster, which leave a steep final entropy profile $K(r) \propto r^{1.3}$.  In Sec.~\ref{results} we incorporate features from all the models discussed above into our set of phenomenological models.\\
\section{SZE Signals from Analytic Models}
\label{results}
In this section we first describe our analytic ICM model taken from \citet{voit/etal:2003} with slight modifications.  Under the assumption of smooth accretion and gravitational shock heating, the mass accretion history determines $K(M_{g})$, which can be modified to account for cooling and a uniform preheating.  Our phenomenological approach generalizes this ICM model to explore parameter ranges suggested or unconstrained by observations.  We no longer assume smooth accretion or attempt to compute $K(M_{g})$ or $f_{\rm ICM}$ from first principles, and we also allow the dark matter potential and boundary pressure to vary.  Because researchers \citep[e.g.,][]{ostriker/bode/babul:2005} sometimes use a polytropic model for the ICM, we investigate its SZE observables and use it to consider a model of lumpy accretion.  Throughout these models we translate the effects of non-gravitational physics and other uncertainties in ICM physics into their effects on the gas entropy profile, cluster potential, and boundary conditions, from which we can easily compute the cluster observables.  Following \citet{voit/etal:2003}, we assume for cosmological parameters $t_{o} = 13.4 \; {\rm Gyr}$, $\Omega_{{\rm m,o}} = 0.3$, $\Omega_{\Lambda{\rm ,o}} = 0.7$, $h = 0.71$, $\sigma_{8} = 0.9$, and the universal baryon fraction $f_{\rm b} = 0.02 h^{-2} \Omega_{{\rm m,o}}^{-1}$.  We do not expect these choices to limit the validity of our results.  In particular, the universal baryon fraction only sets the normalization of the gas density, since we neglect its self-gravitational effects.  We also assume a fully ionized ICM with primordial abundances, so $\rho/n m_{p} = \mu = 0.59$ and $n_{e} = 0.52 n$.  For cooling computations we assume a metallicity $Z = 0.3 Z_{\sun}$.\\
\subsection{Analytic Model of Preheating and Cooling}
\label{hcmodel}
We model the dark matter with an NFW potential \citep{navarro/frenk/white:1997} of overdensity $\Delta(z)$, mass $M_{\Delta}$, and virial radius $r_{\Delta}$:
\begin{equation}
\label{nfwpot}
\phi_{NFW}(r)=\frac{G M_{\Delta}}{r_{\Delta}} \frac{\log (1+c_{\Delta} r/ r_{\Delta})}{\log(1+c_{\Delta})- c_{\Delta}/(1+c_{\Delta})}\frac{ r_{\Delta}}{r}.
\end{equation}
We fix $\Delta(z)$ using the approximation in \citet{bryan/norman:1998} to the spherical top hat collapse model in a $\Lambda CDM$ universe, so $\Delta \approx 100$ at $z=0$.  As in \citet{voit/etal:2003}, we parametrize the concentration parameter's weak dependence on cluster mass and redshift \citep{navarro/frenk/white:1997} as
\begin{equation}
\label{conceqn}
c_{\Delta}(M_{\Delta}) = 8.5 \; (M_{\Delta}/10^{15} h^{-1} M_{\sun})^{-0.086} \; (1+z)^{-0.65},
\end{equation}
and we use the mass accretion histories fit in \citet{voit/etal:2003} for clusters observed at $z=0$ to a merger tree algorithm of \citet{lacey/cole:1993}: 
\begin{equation}
\label{acchist}
\log (M/M_{o}) = a_{1} \log (t/t_{o}) + a_{2} (\log (t/t_{o}))^{2}.
\end{equation}
These functions already introduce some similarity breaking in our baseline model because $a_{1}$, $a_{2}$, and $c_{\Delta}$ all depend on cluster mass.  We consider $10^{15} h^{-1} M_{\sun}$ ($a_{1} = 1.94$ and $a_{2} = -0.55$), $10^{14} h^{-1} M_{\sun}$ ($a_{1} = 1.10$ and $a_{2} = -0.88$), and $10^{13} h^{-1} M_{\sun}$ ($a_{1} = 0.64$ and $a_{2} = -0.96$) clusters at $z=0$ \citep{voit/etal:2003}.\\
In the smooth accretion approximation, infalling gas of uniform density is shocked at a well-defined radius, inside of which we assume hydrostatic equilibrium.  The parameter $\xi$ describes the position of the gas accretion shock radius $r_{\rm ac}$
\begin{equation}
\label{xieqn}
\xi = 1 - \frac{r_{\rm ac}}{r_{\rm ta}}.
\end{equation}
We assume $r_{ta} = 2 r_{\Delta}$ \citep{voit/etal:2003} for the position of the turn-around radius of the gas currently accreting.  We follow \citet{voit/etal:2003} for a self-consistent determination of $\xi(t)$\footnote{We allow their parameter $g_{1}$ to vary and also approximate $r_{\Delta}'(t) = 0.2v_{\rm ac}(t)$.}.  As in \citet{tozzi/norman:2001} and \citet{voit/etal:2003}, $\xi$ is nearly constant in time, except for a sharp jump as accretion goes from adiabatic to shock-dominated in preheating models.  The free-fall velocity of the gas relative to the cluster at the accretion shock is fixed by energy conservation:
\begin{equation}
\label{vaceqn}
v_{\rm ac}^2 = \frac{2GM_{\Delta}\xi}{r_{\rm ac}}.
\end{equation}
Note that we neglect gas energy changes associated with adiabatic compression of the infalling gas and a time-varying amount of enclosed dark matter (see \citet{tozzi/norman:2001} for a discussion of these effects).  We adopt the approximation that dark matter and baryonic mass densities trace one another before accretion:
\begin{equation}
\label{acceqn}
f_{\rm b} M'(t) = 4 \pi r_{\rm ac}^{2} \rho_{1} u_{1}.
\end{equation}
Here $f_{\rm b}$ is the universal baryon mass fraction, $r_{\rm ac} = 2 r_{\Delta} (1-\xi)$ is the gas shock radius, $\rho_{1}$ is the average pre-shock gas density, and $u_{1}$ is the velocity of the gas relative to the moving shock radius.  As our self-consistent determination of $\xi$ suggests, we further assume that the ratio of the shock radius to the dark matter virial radius changes slowly, so that $u_{1}$ is just given by
\begin{equation}
\label{u1eqn}
u_{1} = v_{\rm ac} + 2(1-\xi)r'_{\Delta}(t).
\end{equation}
$\rho_{1}$, $u_{1}$, and the external gas entropy $K_{\rm preheat}$ fix the Mach number of the shock at $r_{\rm ac}$.  We apply the Rankine-Hugonoit jump conditions \citep{landau/lifshitz:FM} to determine the post-shock density and pressure.  Thus $M(t)$ and $\xi(t)$ fix the post-shock entropy profile $K(M_{\rm gas})$.  Note that we do not assume that the post-shock velocity is zero, since the gas is continually compressing to maintain an approximately fixed value of $\xi$.  The post-shock velocity relative to the cluster is $\lesssim 25\%$ of the sound speed throughout the accretion histories, so that hydrostatic equilibrium remains a good approximation.  Our model ignores any subsequent accretion shocks bringing the gas to rest in the cluster; these would occur at higher densities and involve relatively small changes in energy.\\
Once we have the entropy distribution $K(M_{\rm gas})$ that includes preheating and accretion shock heating, we modify $K(M_{\rm gas})$ according to the cooling approximation in \citet{voit/etal:2003}.  Finally we solve the equations of hydrostatic equilibrium and mass conservation to determine the gas property profiles
\begin{eqnarray}
\frac{1}{\rho} \frac{dP}{dr} = -\frac{d\phi_{NFW}}{dr} \nonumber \\
\frac{dM_{\rm gas}}{dr} = 4 \pi r^{2} \rho \label{hydro} \\
P = K(M_{\rm gas}) \rho^{5/3}. \nonumber\\
\nonumber
\end{eqnarray}
We vary $\xi(t_{o})$ until $M_{\rm gas}(r_{\rm ac}) = f_{\rm ICM} M(t_{o})$, where the mass accretion rate, the shock Mach number, and $K(M(t_{o}))$ fix the boundary conditions.  To avoid the singularity associated with $K=0$ when the gas cools, we set a minimum entropy $K_{\rm min} = 0.01 K_{200}$ so that we do not artificially introduce similarity breaking.  SZE cluster properties are unaffected by this choice.\\
To relate our gas profiles to an X-ray observable, we choose to compute $L_{X, \rm cut}$, the X-ray luminosity outside a projected radius $0.05 r_{\Delta}$, to avoid uncertainties about the small entropy values in the model cluster cores.  We also report the total X-ray luminosity, $L_{X}$, and the radius projected onto the sky containing half the total X-ray luminosity, $r_{X}$ (including the region inside $0.05 r_{\Delta}$).  The X-ray luminosity of a small region of gas $dV$ is given by 
\begin{equation}
\label{xraylum}
dL_{X} = n^{2}(x) \Lambda(T) dV.
\end{equation}
$\Lambda(T)$ is the X-ray cooling function, which we approximate with the cooling function that includes all wavelengths given in \citet{sutherland/dopita:1993}.  We compute SZE profiles $y(\theta)$ as well as the SZE luminosity $L_{\rm SZ}$, the central Compton parameter $y_{o}$ ($y$ along the line of sight passing through the cluster center), and the projected radius containing half the SZE luminosity, $r_{\rm SZ}$.\\
\subsection{Preheating and Cooling Model Results}
\label{pcmresults}
\begin{figure}
\includegraphics*[scale=0.85,angle=0]{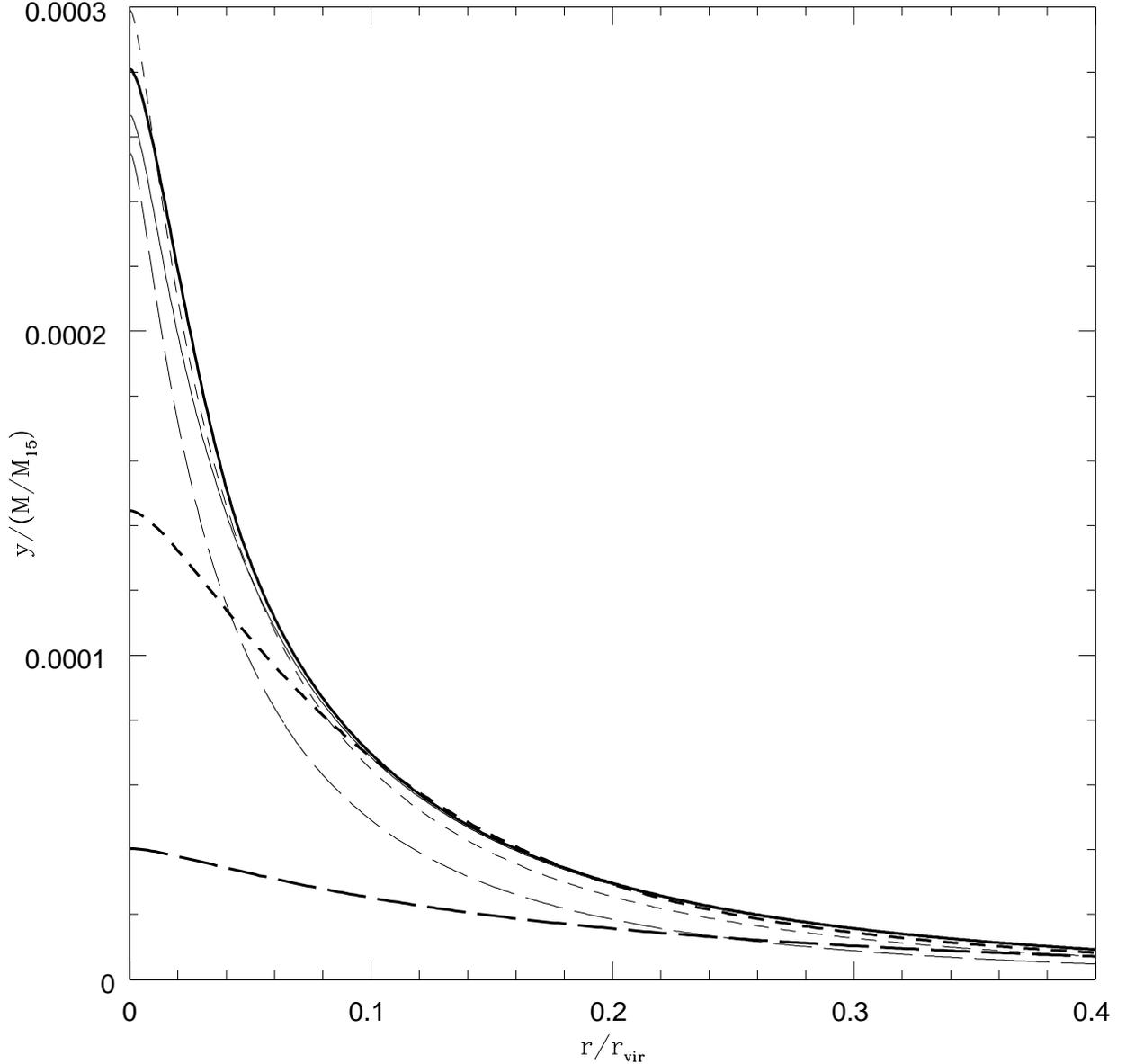}
\caption{\label{fig:szhc}  SZE surface brightness profiles for our cooling only and cooling+preheating ($K_{\rm preheat} =  3 \times 10^{33} {\rm \; erg \; cm^2 \; g^{-5/3}}$) models scaled by $M/10^{15} h^{-1} M_{\sun}$ (see Sec.~\ref{pcmresults} for more discussion).  In a completely self-similar model the profiles of all cluster masses would coincide.  Our cooling model (thin curves) breaks similarity by inducing a mass dependent cooled gas fraction, while our cooling+preheating model (thick curves) produces more diffuse cores in lower mass clusters.  The solid curves are the $10^{15} h^{-1} M_{\sun}$ cluster models.  The primary result of preheating for this cluster is to shut off cooling, so the preheating model has larger gas fraction and thus SZE profile.  Short-dashed curves are the $10^{14} h^{-1} M_{\sun}$ cluster models, where preheating has significantly reduced the central gas density and therefore $y_{o}$.   The long-dashed curves are the $10^{13} h^{-1} M_{\sun}$ cluster models.  The large isentropic core severely extends the gas profile in the preheating model.}
\end{figure}

We first compute the predictions of the model described in Sec.~\ref{hcmodel} when both preheating and cooling are neglected, and then include cooling and allow $K_{\rm preheat}$ to vary in the range  $10^{30} - 10^{34} {\rm \; erg \; cm^2 \; g^{-5/3}}$.   In Appendix A we report X-ray and SZE observable properties of our model clusters.  Our baseline model neglects cooling and sets $K_{\rm preheat} = 0$, and produces an X-ray core surface brightness too large compared to observed cold core regions \citep{markevitch:1998}.  The systematic decrease in $c_{\Delta}$ and increase in accretion pressure with increasing cluster mass already introduce some similarity breaking: low mass clusters are hotter in the cluster interior and have larger accretion radii.  Otherwise gas profiles remain approximately self-similar.  Our baseline model reproduces the self-similar $L_{\rm SZ}$ scaling result, and we find $L_{\rm SZ} = (1.6 \times 10^{-4} \; {\rm Mpc}^{2}) (M/10^{15} h^{-1} M_{\sun})^{5/3}$.\\
With cooling but negligible pre-shock gas entropy, the cooled fraction of baryons was 0.05, 0.15, and 0.42 for the highest to lowest mass clusters.  The cooling approximation of \citet{voit/etal:2003} depends on the assumption of $T = T_{\Delta}$ throughout the cluster at all times, and appears to underestimate the cooled fraction compared to more precise analytic cooling models by around a factor of 2 \citep[see][Fig.~1]{oh/benson:2003}, and more so in cooling-only hydrodynamic simulations \citep{dasilva/etal:2004}.  Cooling steepens the inner entropy profile sufficiently to produce X-ray core surface brightness values in the range measured in the cold core regions of \citet{markevitch:1998}.  Once scaled according to the remaining gas fraction $f_{\rm ICM}$, density and entropy profiles remained approximately self-similar.  The $L_{\rm SZ}-M$ scaling relation slightly steepens to $L_{\rm SZ} = (1.6 \times 10^{-4} \; {\rm Mpc}^{2}) (M/10^{15} h^{-1} M_{\sun})^{1.75}$.  If we scale $L_{\rm SZ}$ by the uncooled gas fraction $f_{\rm ICM}$ as well, we recover the self-similar result found in the no cooling model above: $L_{\rm SZ}/(f_{\rm ICM}/f_{\rm b}) = (1.6 \times 10^{-4} \; {\rm Mpc}^{2}) (M/10^{15} h^{-1} M_{\sun})^{5/3}$.  Thus, in our model the mass-weighted temperature stays fixed when we allow cooling.\\
In addition to cooling we now consider $K_{\rm preheat}$ in the range $10^{30} - 10^{34} {\rm \; erg \; cm^2 \; g^{-5/3}}$.  \citet{voit/etal:2003} adopt $3 \times 10^{33} {\rm \; erg \; cm^2 \; g^{-5/3}}$ to reproduce the $L_{X}-T$ relation, and observations suggest $1.4 \times 10^{33} {\rm \; erg \; cm^2 \; g^{-5/3}}$ \citep{lloyd-davies/ponman/cannon:2000, ponman/cannon/navarro:1999}.  Large $K_{\rm preheat}$ values introduce isentropic, hot cores, particularly in the least massive cluster.  Figure~\ref{fig:szhc} compares the SZE surface brightness profiles produced by the cooling-only and $K_{\rm preheat} = 3 \times 10^{33} {\rm \; erg \; cm^2 \; g^{-5/3}}$ models.  Preheating dramatically broadens SZE profiles in the low mass systems due to the fractionally greater energy input, while the $L_{\rm SZ}$ scaling only slightly steepens for this large $K_{\rm preheat}$ value: $L_{\rm SZ} = 1.6 \times 10^{-4} \; {\rm Mpc}^{2} (M/10^{15} h^{-1} M_{\sun})^{1.70}$.\\
\clearpage
\clearpage
\subsection{Phenomenological Models and Comparison with Observations}
\label{phenomsec}
In this section we adopt a phenomenological approach in order to include information from the latest X-ray observations and to explore a wider range of cluster parameters.  While we still assume spherical symmetry, an NFW potential (Eqn.~\ref{nfwpot}), hydrostatic equilibrium, and a well-defined accretion shock radius bounding the ICM, we do not attempt to compute $K(M_{\rm gas})$ or $f_{\rm ICM}$ from first principles.  Furthermore, we allow the accretion pressure and dark matter concentration to vary.  Motivated by both observations and models discussed in Sec.~\ref{xraysec}, we parametrize the gas entropy by a normalization $K_{\rm max}$ and profile $f(x=r/r_{\Delta})$ with core radius $x_{c}$ and two exponents $s_{1}$ and $s_{2}$:
\begin{equation}
\label{phenent}
f(x) = \max \left(\frac{K_{\rm min}}{K_{\rm max}}, \;
\cases{\left(\frac{x_{c}}{x_{\rm max}}\right)^{s_{2}}  \left(\frac{x}{x_{c}}\right)^{s_{1}} &$x \leq x_{c}$ \cr
\left(\frac{x}{x_{\rm max}}\right)^{s_{2}} &$x \geq x_{c}$ \cr}
\right).
\end{equation}
A power-law parametrization is consistent with both the predictions from gravitational heating and quasar blasts and with X-ray observations of entropy profiles (see Sec.~\ref{xraysec}).  A flatter entropy profile ($s_{1} < s_{2}$) in the core is consistent with observations \citep{pratt/arnaud:2003,ponman/sanderson/finoguenov:2003} and preheating models.\\
Solving the hydrostatic equilibrium equation for the normalized pressure profile $p(x) = P(x)/P(r_{\rm ac})$ with $x = r/r_{\Delta}$ we find
\begin{equation}
\label{phenomsoln}
p(x)^{2/5}-1 = \frac{2}{5}\frac{\mu m_{p}}{T_{\rm max}} \int_{x}^{x_{\rm max}} \frac{d \phi_{NFW}}{dy}f(y)^{-3/5} dy\\
\end{equation}
with $T_{\rm max} = \mu m_{p} P(r_{\rm ac})^{2/5} K_{\rm max}^{3/5}$, the gas temperature at $r_{\rm ac}$.  We fix $P(r_{\rm ac})$ by the gas accretion rate $f_{\rm ICM} M'(t)$, incoming velocity $u_{1}$ (Eqn.~\ref{u1eqn}), and assuming a strong shock at $t_{o}$ (that is, the external thermal pressure can be neglected at $t_{o}$).  We vary $r_{\rm ac}$ until the enclosed gas mass is $M_{\rm gas} = f_{\rm ICM} M_{\Delta}$.  We do not assume smooth accretion, so $K_{\rm max}$ varies independently of $P(r_{\rm ac})$.  Note that for fixed $T_{\rm max}$, the pressure, density, and SZE profiles are fixed, and their normalizations simply scale with $f_{\rm ICM}$.  This scaling will be broken only if cluster physics condenses or ejects a different fraction of gas compared with the local infalling average.  A fit to the accretion rates at $t_{o}$ from \citet{voit/etal:2003} yields $M'(t_{o}) = (1.94 \times 10^{15} h^{-1} M_{\sun} t_{o}^{-1}) ((M(t_{o})/10^{15} h^{-1} M_{\sun})^{5/4} w_{\rm accr}$, where we allow a ``fudge factor'' $w_{\rm accr}$ in the accretion rate.\\
In our ``self-similar'' model we set $K_{\rm max} = K_{100}$, $K_{\rm min} = 0.01 K_{100}$, $s_{1} = s_{2} = 1.1$, $x_{c}$ = 0, $c_{\Delta}$ according to Eqn.~\ref{conceqn}, and $f_{\rm ICM} = 0.13$.  These are the expected values in the gravitational-only heating scenario described in Sec.~\ref{xraysec} (recall $\Delta(z=0) = 100$ in $\Lambda CDM$), with further cooling and condensation of baryons in clusters neglected.  Parameters in other models are fixed to these values unless specified otherwise.  We list the computed observables from our models in Appendix A.  Figure~\ref{fig:figd15} and Figure~\ref{fig:figt15} show the density and temperature profiles for phenomenological models of the most massive cluster; the others are similar.\\
According to $\Lambda CDM$ numerical simulations \citep{dolag/etal:2004}, dark matter concentration parameters show intrinsic scatter in a log-normal distribution with $\ln (c/c_{avg}) \approx 0.22$, and a range of mean values have been reported \citep{navarro/frenk/white:1997,komatsu/seljak:2001,dolag/etal:2004}.  We allow $c_{\Delta}$ to range by 8 to encompass these uncertainties, and find a $\approx 20\%$ decrease in $L_{\rm SZ}$ for the lowest values of $c_{\Delta}$ and $\approx 10\%$ increase for the highest for all cluster masses considered.  Taking the mean $c_{\Delta}$ and its scatter reported by \citet{dolag/etal:2004} and our self-similar gas model, we find that the intrinsic scatter in $c_{\Delta}$ induces variations in $L_{SZ}$ of $\lesssim 8\%$.  The effect is smaller at lower concentrations (higher mass clusters) and shallower entropy profiles (lower $s_{1}$).  Figure~\ref{fig:figd15} and Figure~\ref{fig:figt15} show that increasing $c_{\Delta}$ both concentrates more gas in the inner, hotter cluster region and significantly increases the temperature in that region.\\
%
%
Mass accretion is stochastic, aspherical, and has not been directly observed at the cluster boundary.  We vary the accretion pressure by a factor $w_{\rm accr} = 3.5$ and $w_{\rm accr} = 1/3.5$.  The additional accretion pressure boosts $L_{\rm SZ}$ by $44\%$ in our largest mass cluster and pushes the gas temperature well above $T_{\Delta}$ at $r_{\rm ac}$ (see Figure~\ref{fig:figt15}).  All other cases saw much milder changes in $L_{\rm SZ}$.  For the $10^{14} h^{-1} M_{\sun}$ and $10^{13} h^{-1} M_{\sun}$ clusters we also list the results that would be obtained in the case of a self-similar mass accretion rate ($M'(t_{o}) \propto M(t_{o})$).\\
We allow for a core in the entropy profile, as suggested by preheating models and observed profiles, but assume a power-law entropy profile outside the core.  The entropy profile is observationally unconstrained near the cluster virial radius, but gravitational heating probably dominates in that region.  We allow the power law index to range from 0.7 to 1.5, incorporating the observed $K(r) \propto r^{0.94 \pm 0.14}$ \citep{pratt/arnaud:2005}, $K(r) \propto r^{1.1}$ expected from gravitational heating \citep{tozzi/norman:2001,borgani/etal:2001b},  and the steep profile $K(r) \propto r^{1.3}$ expected from quasar blasts \citep{lapi/cavaliere/menci:2005}.  We vary $K_{max}$ over more than a factor of 3, which should bound the expected range for clusters between $10^{13} h^{-1} M_{\sun}$ and $10^{15} h^{-1} M_{\sun}$.  Variations on $s_{2}$, inclusion of an entropy core, fixing $K_{\rm min} = 3.5 \times 10^{33} {\rm \; erg \; cm^2 \; g^{-5/3}}$, and varying $K_{\rm max}$ between $K_{200}$ and $3 K_{100}$ all caused $\leq 5\%$ variation of $L_{\rm SZ}$ for our $M = 10^{15} h^{-1} M_{\sun}$ cluster.  Even though increasing $K_{\rm max}$ raises the temperature profiles, the gas also becomes more extended and occupies the cooler regions of the cluster.  Our $10^{14} h^{-1} M_{\sun}$ and $10^{13} h^{-1} M_{\sun}$ clusters were slightly more sensitive to changes in the entropy normalization, but $L_{\rm SZ}$ still remained within $\approx 10\%$ of the self-similar model.\\
We consider two scenarios consistent with the observationally suggested $K_{\rm max} \propto T_{\Delta}^{2/3}$ \citep{ponman/sanderson/finoguenov:2003}.  If the primary effect of preheating is to reduce the pre-shock gas density by a factor dependent on cluster potential depth \citep{voit/etal:2003, ponman/sanderson/finoguenov:2003,voit/bryan:2005} or if the fractional feedback entropy injection into the bound gas depends on cluster mass, then we expect the entropy profile normalization to deviate from the self-similar scaling while $f_{\rm ICM}$ remains independent of cluster mass.  If instead the similarity-breaking is introduced because cooling and AGN/supernovae feedback produce a mass-dependent ICM mass fraction, then $K_{\rm max} \propto T_{\Delta}^{2/3}$ implies $f_{\rm ICM} \propto T_{\Delta}^{1/2}$.  Changing the entropy normalization with fixed $f_{\rm ICM}$ changes $L_{\rm SZ}$ by $\approx 10\%$ in the $M = 10^{13} h^{-1} M_{\sun}$ cluster (where $K_{max}$ is increased by a factor of 2.78).  Introducing a mass-dependent baryon fraction to account for this entropy scaling is equivalent to simply scaling the density profile and thus $L_{\rm SZ}$ by $f_{\rm ICM}/f_{\rm b}$.  Note that this model implies an unreasonable reduction in the ICM mass fraction by 79\% for the $10^{13} h^{-1} M_{\sun}$ cluster.\\
For completeness we also considered the ``threshold'' cooling model from \citet{voit/etal:2002}.  This model estimates the threshold entropy below which gas would condense by the current age of the universe, removes all gas below the threshold, and leaves the entropy profile for the remaining gas unmodified.  The authors suggest that profile describes either strong feedback that ejects the gas below the entropy threshold to well beyond the virial radius, or weak feedback where the gas simply cools.  Cooling of the gas above the entropy threshold is also neglected.  For the $M = 10^{15} h^{-1} M_{\sun}$ ($K_{\rm max} = 2.5 K_{100}$) and $M = 10^{14} h^{-1} M_{\sun}$ ($K_{\rm max} = 1.5 K_{100}$) clusters, this model yielded low $L_{\rm SZ}$ values, but once accounted for the missing gas fraction, the $L_{\rm SZ}$ values were within $2\%$ of the self-similar value.  This reinforces our result that variation in the entropy profiles affect the average gas temperature only very weakly.  The prescribed method in \citet{voit/etal:2002} for computing the unmodified entropy distribution yielded a non-monotonic entropy profile for the $ M = 10^{13} h^{-1} M_{\sun}$ cluster, and so we did not complete the calculation for this cluster.\\
%
%
In Figures~\ref{fig:figdobs} - \ref{fig:figtobs} we compare our phenomenological density and temperature profiles with 12 nearby, relaxed clusters from \citet{vikhlinin/etal:2005}.  We use the $10^{15} h^{-1} M_{\sun}$ cluster values, which are in the best agreement with the NFW fit values in \citet{vikhlinin/etal:2005}, $c_{100} = 3.6 - 9.2$.  Our adopted value $f_{\rm b} = 0.132$ is $\approx 25\%$ below the universal value constrained by CMB observations to $0.175 \pm 0.023$ \citep{readhead/etal:2004b,spergel/etal:2003} and is in reasonable agreement with the observed stellar component of clusters \citep{roussel/sadat/blanchard:2000, balogh/etal:2001, lin/mohr/stanford:2003}.  Near $r_{500}$, observed density and temperature profiles agree with the model curves with $w_{\rm accr} = 1$ (see Figures \ref{fig:figt15}, \ref{fig:figdobs}, and \ref{fig:figtobs}), consistent with our assumptions about the outer gas, accretion pressure, and $f_{\rm b}$.  In Figure~\ref{fig:figdobs}, we also see the $T > 5 \; {\rm keV}$ observed cluster density profiles suggest an entropy profile shallower than $r^{1.1}$ and/or lower $c_{\Delta}$, particularly in the core.  The average temperature profile agrees well with our $r^{0.7}$ model (long dashed curve in Figure~\ref{fig:figtobs}) outside $0.2 r_{100}$.  This comparison indicates that our chosen parameter values cover a range of gas profiles at least as large as the scatter in the observations.\\
Under the self-similarity assumption, $\rho(r/r_{\rm vir})$ is independent of cluster mass.  However, densities in the $T < 5 \; {\rm keV}$ clusters (see Figure~\ref{fig:figdobs2}) are significantly lower than in the hottest clusters (Figure~\ref{fig:figdobs}).  We devise two models to trace the lowest density profile in Figure~\ref{fig:figdobs2}.  The resulting temperature profiles (thick long-short dashed curve in Figure~\ref{fig:figtobs}) agrees with the observed hot cluster average profile out to $0.35 r_{vir}$, and has a sharper inner peak, as observed in \citet{vikhlinin/etal:2005} cool clusters.  Fits 1 and 2 both have $s_{1} = s_{2} = 0.7$ and $c_{\Delta} = 7.7$, the value of $c_{100}$ expected by \citet{dolag/etal:2004} for a $10^{14} h^{-1} M_{\sun}$ cluster.  In fit 1 we keep the same ICM mass fraction, $f_{\rm ICM} = 0.13$, as in the phenomenological models shown in Figures~\ref{fig:figd15} - \ref{fig:figdobs}, and raise the gas entropy normalization to $K_{max} = 2.5 K_{100}$.  In contrast, fit 2 assumes the accretion pressure is set by gas with $f_{\rm b} = 0.13$, while the ICM mass fraction inside the cluster is $f_{\rm ICM} = 0.5 f_{\rm b}$, either due to ejection or increased condensation.  The entropy normalization in fit 2 is the self-similar value for this ICM mass fraction: $K_{max} = K_{100} (0.5 f_{\rm b})^{-2/3}$.  By design, fits 1 and 2 are nearly indistinguishable in the X-ray observable region ($r \lesssim r_{500}$ both in density and temperature profiles).  However, fit 1 extends to $1.56 r_{\rm vir}$ and contains all of the gas initially associated with the region, while fit 2 extends only to $0.84 r_{\rm vir}$ and contains half as much hot gas.  This example demonstrates the inability of X-ray observations to constrain the distribution of non-gravitational heating required for similarity breaking; in fit 1, the non-gravitational heating increases the gas entropy levels and its potential energy, while in fit 2, the non-gravitational heating ejects half of the gas from the cluster region and leaves the remaining gas undisturbed.  However, the SZE observables for these two models are different, and we shall return to this point in Sec.~\ref{fbdiscussion}.\\
\clearpage
\begin{figure}
\includegraphics*[scale=0.85,angle=0]{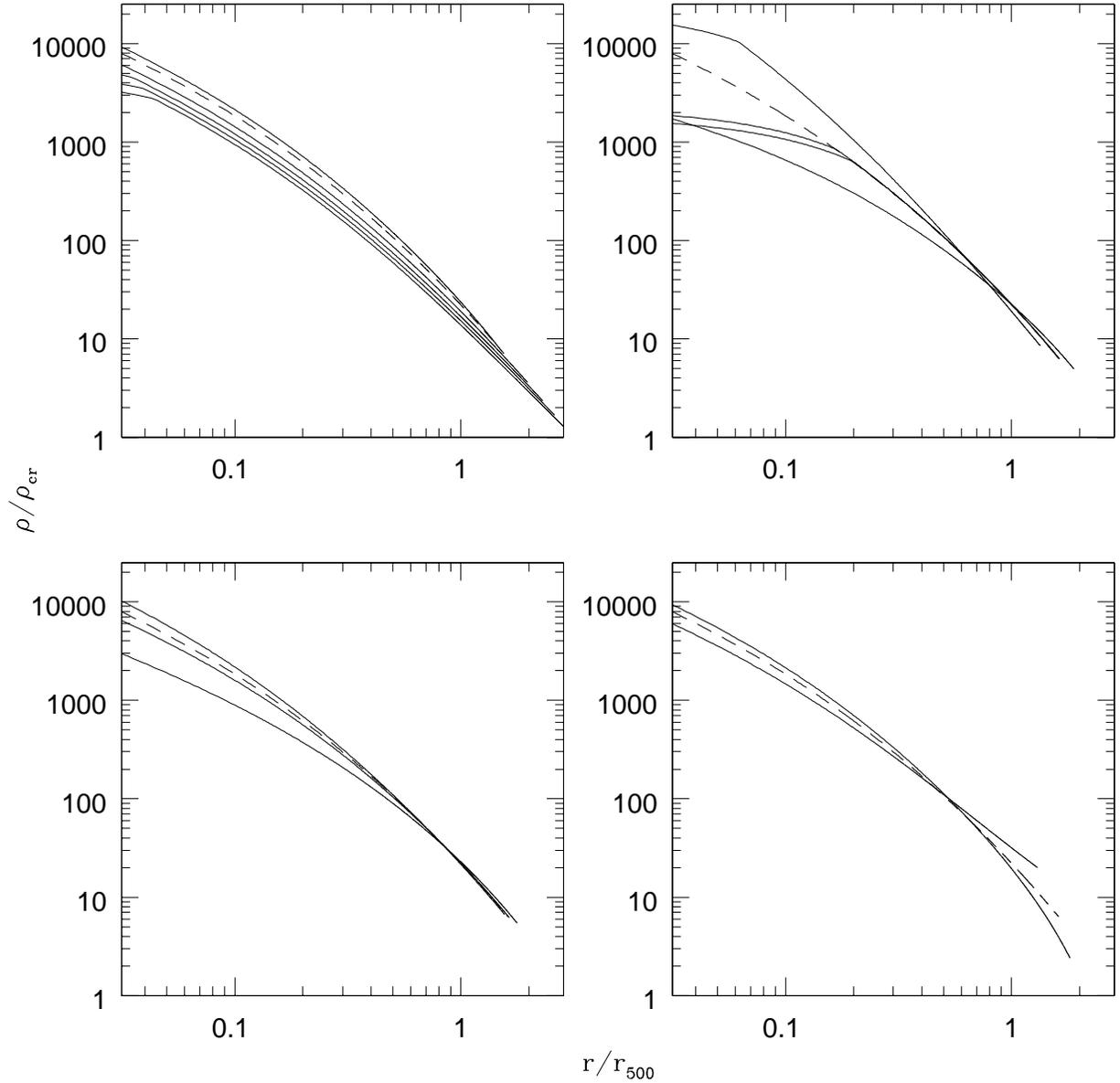}
\caption{\label{fig:figd15}  Phenomenological model density profiles for our $10^{15} h^{-1} M_{\sun}$ cluster.  See Table~\ref{table15} for corresponding observables.  The self-similar model is shown in each panel as the dashed curve.  In the upper left $K_{\rm max}$ is varied between $K_{200}$ and $3K_{100}$.  The central density decreases and $r_{\rm ac}$ increases as $K_{\rm max}$ increases.  In the upper right the entropy profile is varied.  The central density increases with radial entropy power $s_{2}$ (outer curves), but flattens when an entropy core is included (inner curves).  In the lower left the central density increases with $c_{\Delta}$ for $c_{\Delta} = 3$ to $c_{\Delta} = 11$.  In the lower right the accretion pressure is varied through $w_{accr}$.  The density at $r_{\rm ac}$ increases with $w_{\rm accr}$, while the central density decreases.}
\end{figure}
%
%
\begin{figure}
\includegraphics*[scale=0.85,angle=0]{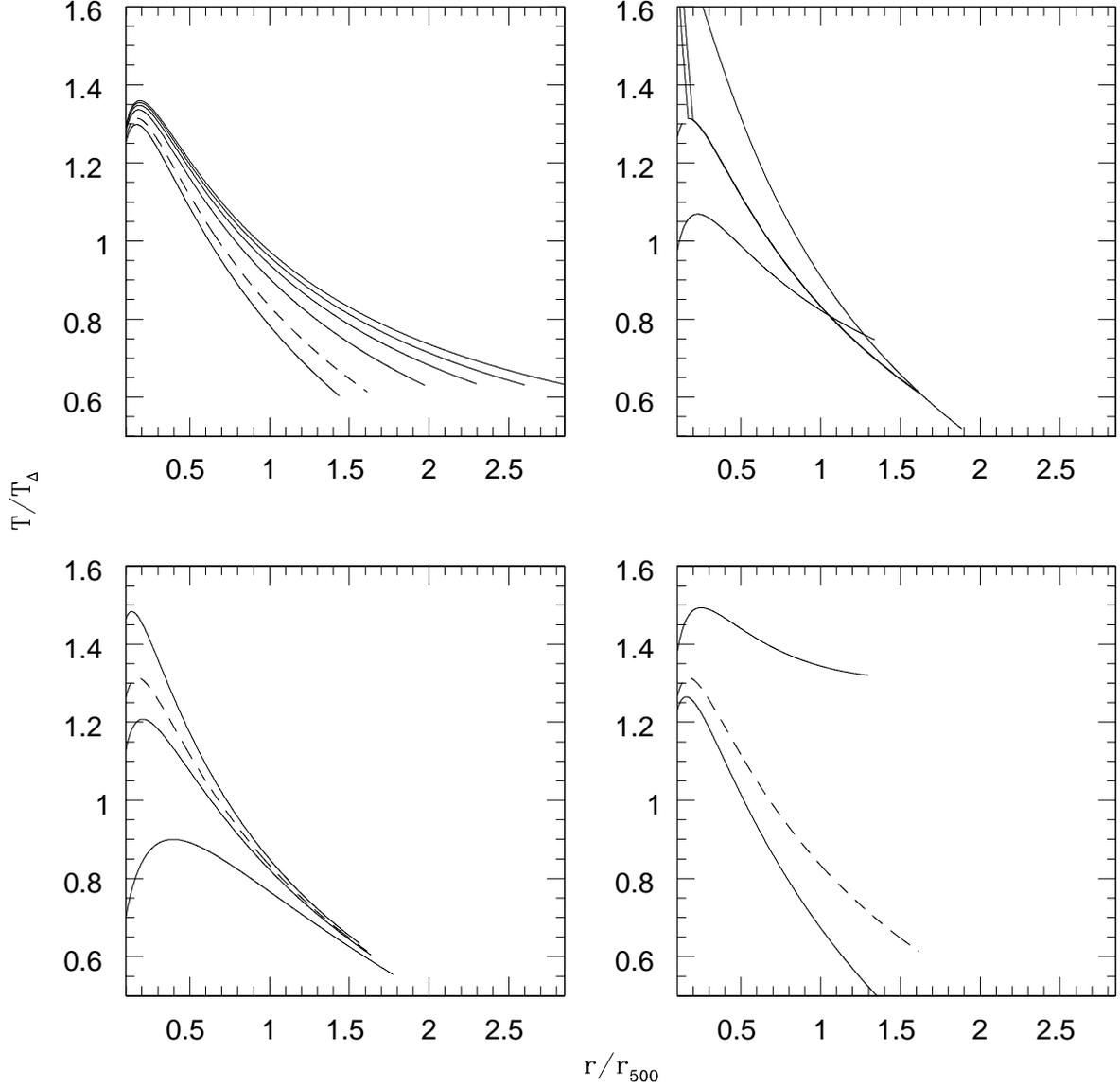}
\caption{\label{fig:figt15}  Phenomenological model temperature profiles for our $10^{15} h^{-1} M_{\sun}$ cluster scaled by the dark matter virial temperature $T_{\Delta}$.  Panels as in Figure~\ref{fig:figd15}.  At fixed radius temperature increases with $K_{\rm max}$ (upper left), $c_{\Delta}$ (lower left), and $w_{\rm accr}$ (lower right).  Temperature decreases with radial entropy power $s_{2}$ (outer curves in the upper right), and an entropy core produces sharply peaked central temperature profiles (upper right).}
\end{figure}
\begin{figure}
\includegraphics*[scale=0.85,angle=0]{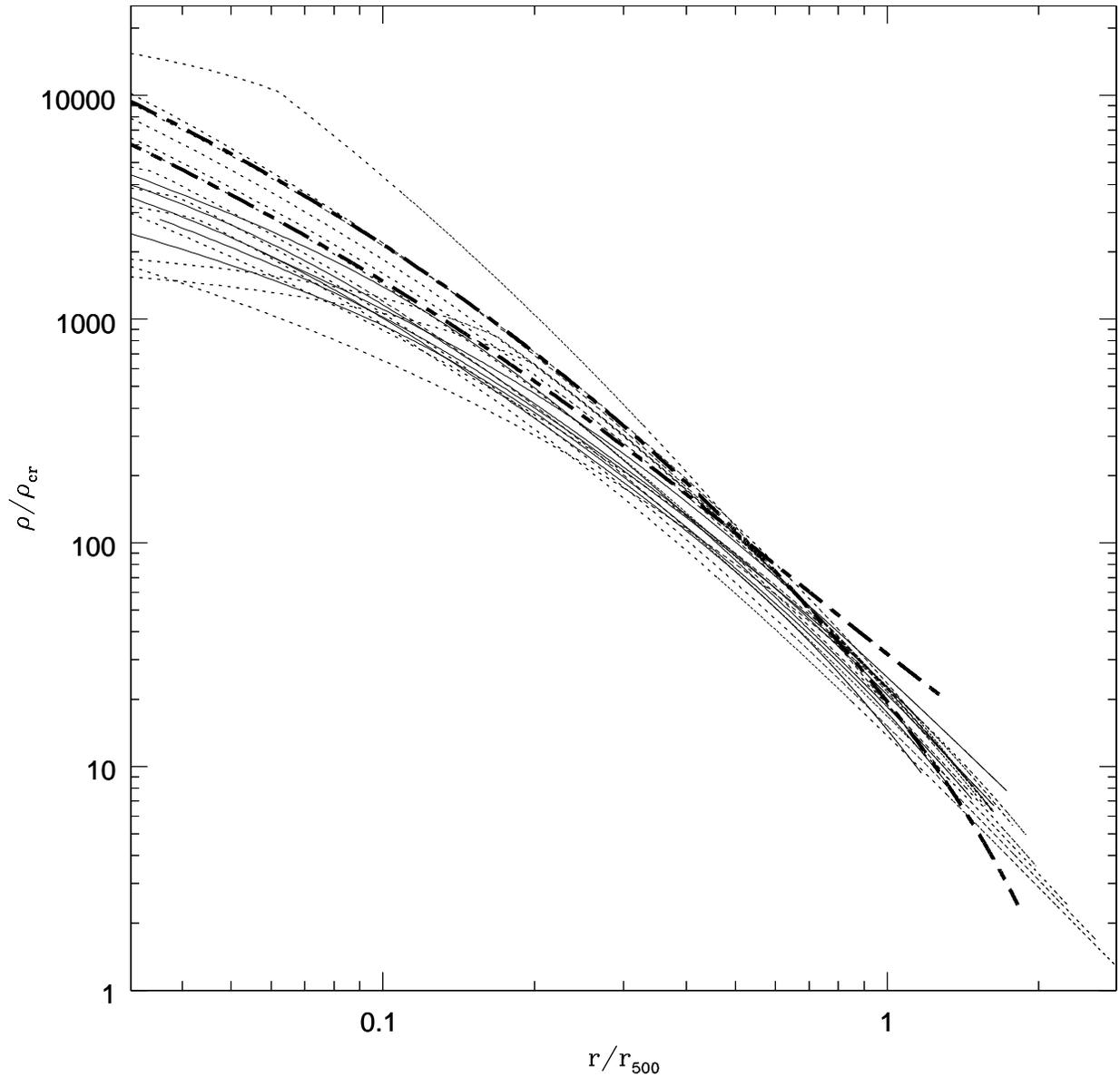}
\caption{\label{fig:figdobs}  All phenomenological model density profiles for our $10^{15} h^{-1} M_{\sun}$ clusters (dotted curves, also in Figure~\ref{fig:figd15}) compared to observed $T > 5 \; {\rm keV}$ clusters from \citet{vikhlinin/etal:2005} (solid curves).  The observed density profiles are consistent with $w_{accr} = 1$; our models varying the accretion pressure (thick long-short dashed curves) do not agree the the observed density gradient at large radii.  The observed profiles are consistent with models with lower $c_{\Delta}$ or more entropy at small radii compared with our self-similar model.}
\end{figure}
\begin{figure}
\includegraphics*[scale=0.85,angle=0]{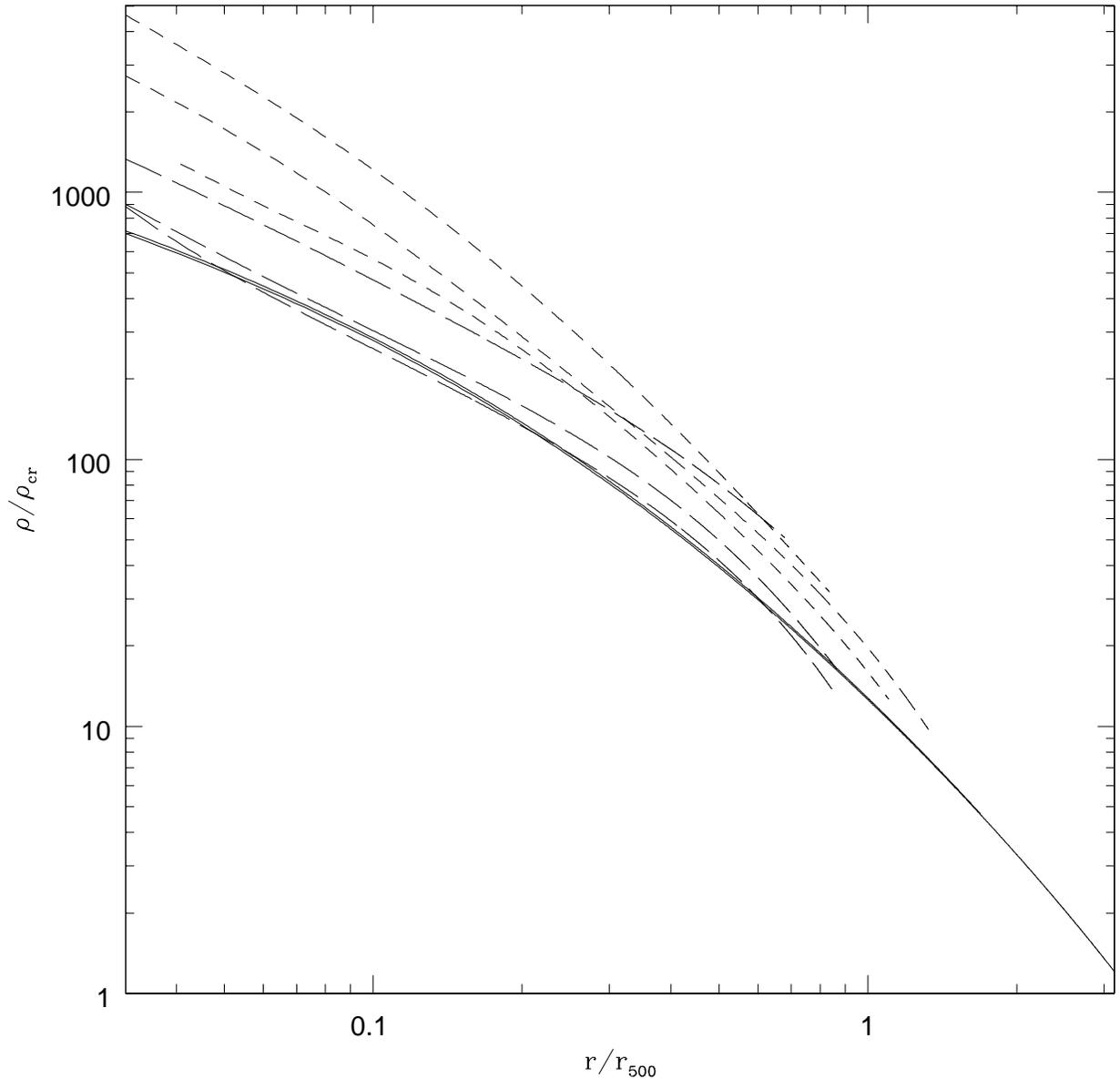}
\caption{\label{fig:figdobs2}  Observed cool clusters from \citet{vikhlinin/etal:2005} grouped by temperature: $2.5 \; {\rm keV} < T < 5 \; {\rm keV}$ (short dashed curves), and $T < 2.5 \; {\rm keV}$ (long dashed curves).  The solid curves are two additional phenomenological models devised to trace the lowest density profiles in the observed region and discussed in Sec.~\ref{phenomsec}.  While the accreting gas is assumed to have the same baryon fraction in both models, one has 50\% less gas in the ICM.  These model density profiles are indistinguishable out to the maximum observed radius, though $r_{\rm ac}$ is much lower in the $f_{\rm ICM} = 0.5 f_{\rm b}$ model.  These models demonstrate the degeneracy between a high entropy level and low baryon fraction in the region of gas currently X-ray observable.}
\end{figure}
\begin{figure}
\includegraphics*[scale=0.85,angle=0]{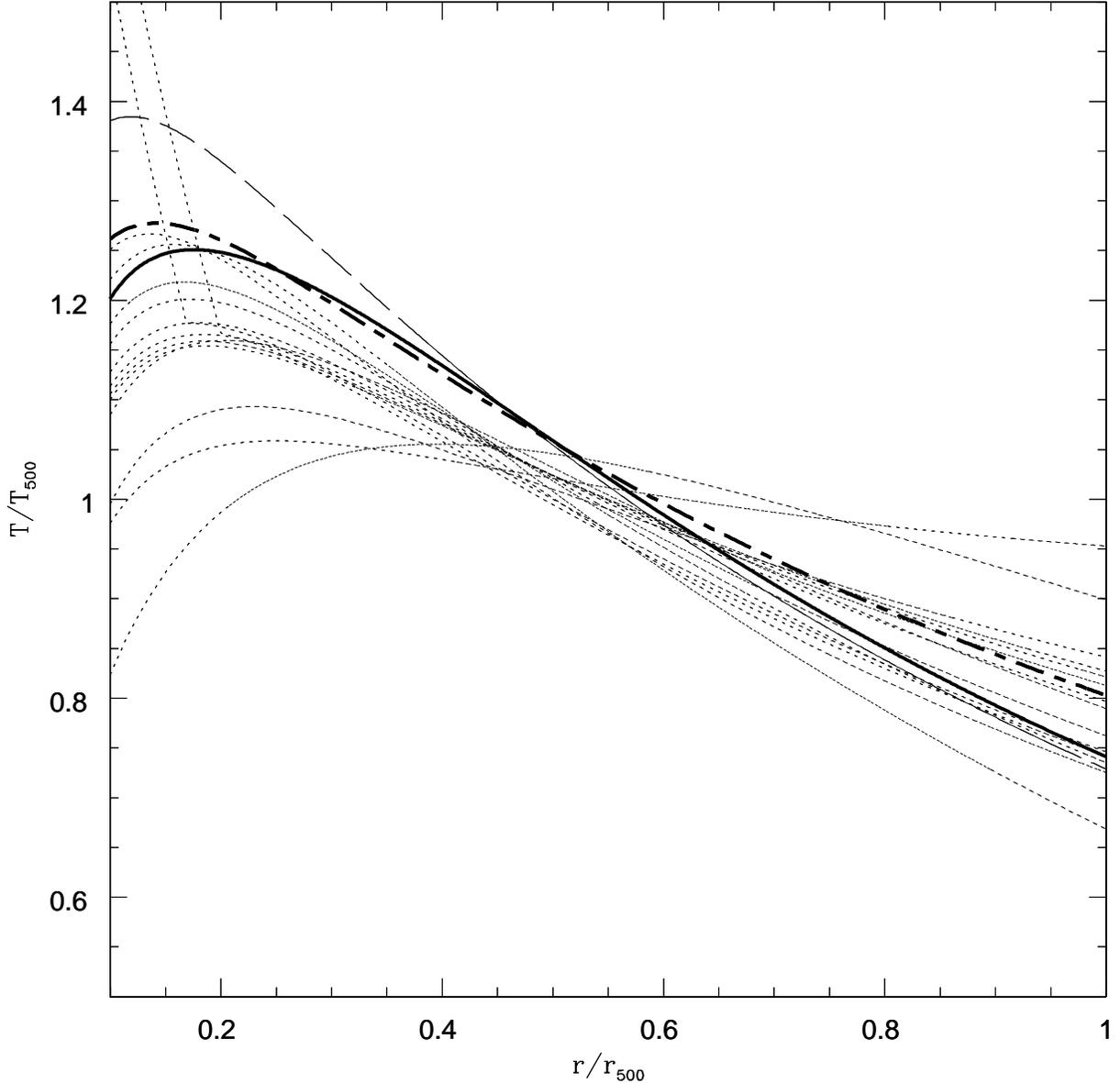}
\caption{\label{fig:figtobs}  Temperature profiles for our $10^{15} h^{-1} M_{\sun}$ model clusters (dotted curves) as in Fig.~\ref{fig:figt15}, but each normalized by their gas mass-weighted temperature within $r_{500}$, $T_{500}$.  The thick solid line is an average temperature profile for observed clusters with $T > 2.5 \; {\rm keV}$ from \citet{vikhlinin/etal:2005}.  As can be seen from Fig.~\ref{fig:figt15}, the gradient of $T/T_{500}$ increases with increasing $c_{\Delta}$ and decreasing $K_{\rm max}$, $w_{accr}$, and $s_{2}$.  Outside $0.2 r_{\rm vir}$ our $s_{2} = 0.7$ model agrees well with the average observed profile (long dashed curve).  The thick short-long dashed curve is the temperature profile of our two additional phenomenological models (indistinguishable on this plot) devised to fit low temperature cluster density profiles (see Sec.~\ref{phenomsec}). They provide a good fit to the average temperature profile out to $\approx 0.35 r_{\rm vir}$.}
\end{figure}
\clearpage
Finally we consider a simple polytropic model $P \propto \rho^{\gamma}$ in an NFW potential and fix $\gamma = 1.2$, in rough agreement with both simulations and observations, at least outside the core (see \citet{ostriker/bode/babul:2005}, Appendix A of \citet{voit/etal:2003}, and references therein).  The hydrostatic equilibrium equation then reduces to 
\begin{equation}
\label{poly}
\frac{dT}{dr} = -\mu m_{p} \frac{\gamma - 1}{\gamma} \frac{d\phi_{NFW}}{dr}.
\end{equation}
The NFW concentration parameters are fixed again by Eqn.~\ref{conceqn}.  The free parameters in the model are the constant of integration in the temperature profile obtained from Eqn.~\ref{poly}, the normalization of the density profile, and the accretion radius.  We assume a strong shock at the accretion radius and set the pressure at the accretion radius as in the models above.  Requiring $f_{\rm b} M(t_{o})$ gas to be contained within the accretion radius constrains the density normalization.  However, we do not wish to impose a strong condition on the density at the accretion radius, since the global density of the gas accreting at $t_{o}$ is not the density determining the post-shock entropy in lumpy accretion; compression of gas accreting in distinct, dense lumps will not significantly raise the global density at $r_{\rm ac}$.  We vary $x_{\rm ac}$, and the other free parameters are then determined by the conditions above.  We exclude models where the post-shock density does not fall between $\rho_{1}$ and $4 \rho_{1}$, the strong shock smooth accretion limit.  Note that these models produce entropy profiles with $K(0) > 0$ and monotonically decreasing temperature profiles (see Eqn.~\ref{poly}).  Again as shown in Appendix A we find the $L_{\rm SZ}$ signal to be extremely robust to our variation of $x_{\rm ac}$ (and thereby the post-shock entropy normalization).\\
\section{Scaling Relations}
\label{relations}
\subsection{Scalings between Observables}
\begin{figure}
\includegraphics*[scale=0.85,angle=0]{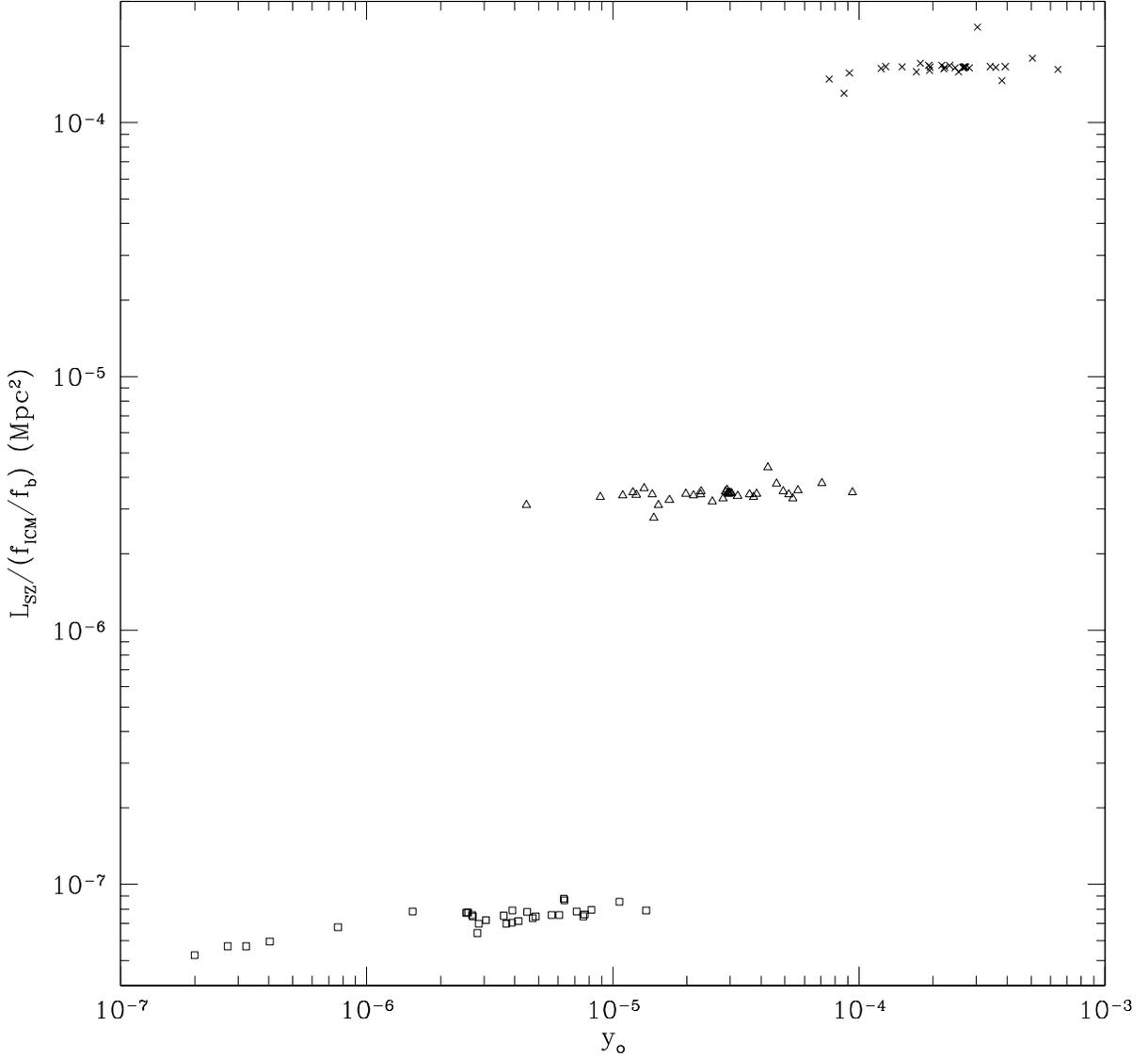}
\caption{\label{fig:SZ1}  Results from our analytic baseline, preheating, and cooling models with $K_{\rm preheat}$ ranging from $10^{30}$ to $10^{34} {\rm \; erg \; cm^2 \; g^{-5/3}}$, as well as all our phenomenological and polytropic models (see Sec.~\ref{results}) varying the entropy profile $K(r)$, dark matter concentration $c_{\Delta}$, accretion pressure, and ICM mass fraction.  Crosses - $10^{15} h^{-1} M_{\sun}$ cluster models, triangles - $10^{14} h^{-1} M_{\sun}$ cluster models, squares - $10^{13} h^{-1} M_{\sun}$ cluster models.  The SZE luminosity, $L_{\rm SZ}$, normalized by $f_{\rm ICM}/f_{\rm b}$, shows little variation within our class of models.  Thus, given the ICM mass fraction $f_{\rm ICM}$, $L_{\rm SZ}$ is an excellent cluster mass indicator.  The central Compton parameter, $y_{o}$, varies widely in our class of models and thus contains information on the thermal history of the gas.}
\end{figure}

While our set of models show a large variation in X-ray properties and central Compton parameter (see Appendix A), the total SZE luminosity is remarkably robust to the variations introduced in our models.  In Figure~\ref{fig:SZ1} we see that the scatter for our model clusters is small in $L_{\rm SZ}/(f_{\rm ICM}/f_{\rm b})$, indicating a tight $L_{\rm SZ}-M_{\Delta}$ relationship for known $f_{\rm ICM}$, despite the consideration of a wide range of possible entropy profiles and cluster unknowns.  Note that $L_{\rm SZ} \propto f_{\rm ICM}$ makes the relation sensitive to the fraction of baryons cooled or ejected from the ICM.  This plot corroborates the finding in \citet{mccarthy/etal:2003a} that $y_{o}$ is much more sensitive to $K_{\rm preheat}$ (or other non-gravitational physics) than is $L_{\rm SZ}$.\\
\clearpage
\begin{figure}
\includegraphics*[scale=0.75,angle=0]{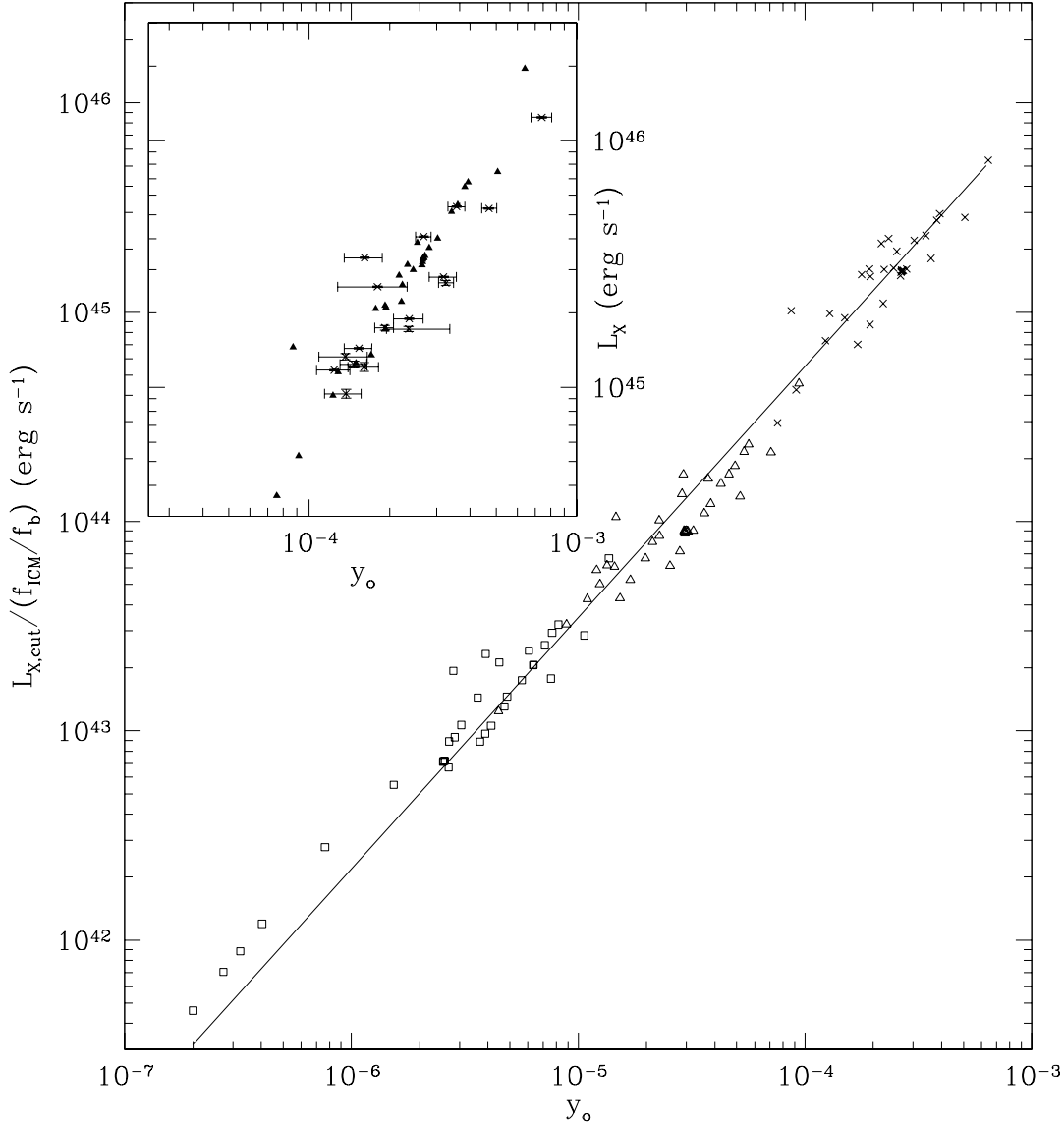}
\caption{\label{fig:XSZ}  $L_{X,cut}$, the X-ray luminosity outside projected radius $0.05 r_{\Delta}$, vs. central Compton parameter, $y_{o}$.  Crosses - $10^{15} h^{-1} M_{\sun}$ cluster models, triangles - $10^{14} h^{-1} M_{\sun}$ cluster models, squares - $10^{13} h^{-1} M_{\sun}$ cluster models.  The power law fit shown has exponent $1.2$.  Note that clusters spanning two orders of magnitude in mass and a wide variety of cluster parameters are well described by a single $y_{o}-L_{X,cut}$ relation.  In the inset we compare the observations assembled in \citet{mccarthy/etal:2003b} (crosses with error bars) to our $10^{15} h^{-1} M_{\sun}$ cluster models (triangles).  The $y_{o}-L_{X}$ normalization depends on mass for our models and is steeper than $y_{o}-L_{X,cut}$.  The observed clusters are very massive, and we find good agreement between the observations and our $10^{15} h^{-1} M_{\sun}$ model clusters.}
\end{figure}
\clearpage
For the class of models studied here we also find a tight correlation between the X-ray luminosity outside the core, $L_{X, \rm cut}$, and the central Compton parameter (see Figure ~\ref{fig:XSZ}).  This is not surprising if $n_{\rm ICM}(r) \sim 1/r^{2}$ outside the core, so that if we ignore any radial temperature dependence, $y \sim \int n_{\rm ICM} dr$ and $L_{x} \sim \int n^{2}_{\rm ICM} r^{2} dr$ are proportional to one another.  Our scaling $L_{X} \propto y_{o}^{1.2}$ is in excellent agreement with similarly robust relations reported by \citet{cavaliere/menci:2001} and \citet{mccarthy/etal:2003a}.  We find the relation is closer to mass-independent when excluding the core X-ray luminosity, but we demonstrate agreement between our model and observations assembled by \citet{mccarthy/etal:2003b} in the $y_{o}-L_{X}$ plane.\\
\subsection{Cluster Energetics}
\label{energetics}
\begin{figure}
\includegraphics*[scale=0.85,angle=0]{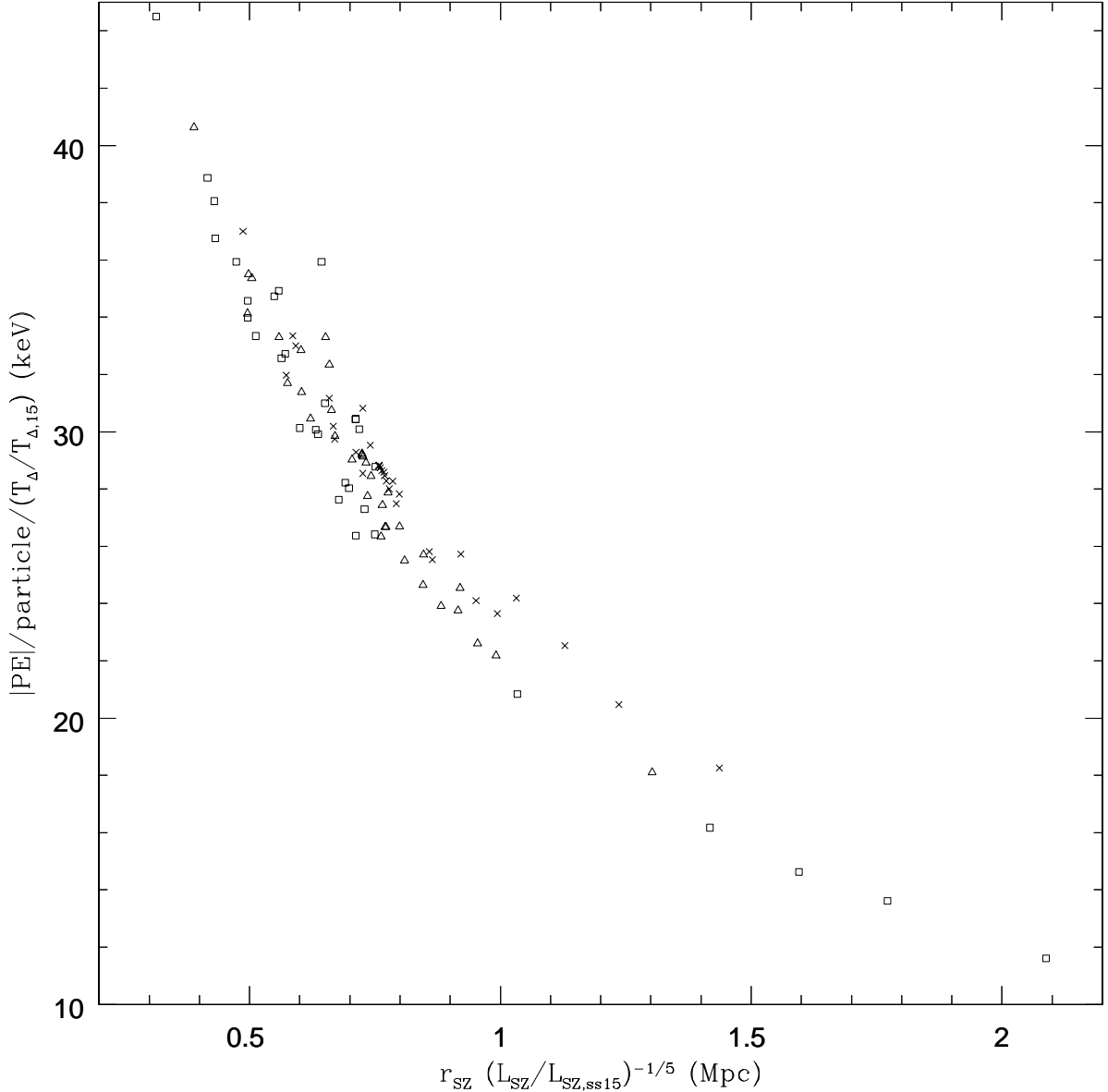}
\caption{\label{fig:potential}  The potential energy per particle scaled by the cluster virial temperature $T_{\Delta}$ can be estimated by the SZE observable $r_{\rm SZ} L_{\rm SZ}^{-1/5}$.  This relation holds over a mass range of 2 orders of magnitude and over our entire range of model variations.  $T_{\Delta}$ is fixed by the cluster mass and so is also tightly related to $L_{\rm SZ}$.  Thus the total potential energy per particle can be estimated from an SZE observation measuring $r_{\rm SZ}$ and $L_{\rm SZ}$.  Crosses - $10^{15} h^{-1} M_{\sun}$ cluster models, triangles - $10^{14} h^{-1} M_{\sun}$ cluster models, squares - $10^{13} h^{-1} M_{\sun}$ cluster models.  $T_{\Delta,15}$ is the virial temperature and $L_{SZ,ss15}$ is the self-similar SZE luminosity for a $10^{15} h^{-1} M_{\sun}$ cluster.}
\end{figure}

Thus far we have focused on the effects of the gas entropy profile on ICM observables.  In an attempt to understand the stability of our $L_{\rm SZ}$ signal to such wide variations, we compute changes in potential and thermal energies induced by changes in model parameters.  To compute the gas potential energy, we estimate the cluster potential by an NFW profile, thus ignoring the deviation of the gas density profile from the dark matter profile:
\begin{equation}
PE_{\rm gas} = \int \rho_{\rm ICM} \phi_{NFW} dV.
\end{equation}
The gas thermal energy is directly proportional to $L_{\rm SZ}$.  Appendix A shows that changes in total gas energy result mostly in changes in potential energy.  Moreover, fractional deviations of the mean energies of our models are on average larger in potential than thermal.  Thus, even with uncertainty about the amount of energy injected into the ICM throughout its history, for the class of models studied here most of the injected energy manifests itself in potential energy.  Note that our models produce total energy values differing by at least a factor of two for a given cluster mass.\\
Consider the two model variations inducing the greatest $L_{\rm SZ}$ signal variation.  Increasing the dark matter concentration parameter effectively increases the temperature associated with the dark matter potential, and so in that case we expect a significant change in thermal energy if $T_{\rm gas} \sim T_{dark}$.  Similarly, an increase in external pressure requires an increase in the internal pressure, thus driving up the thermal energy of the outer gas.\\
An SZE-only observation can infer both the thermal ($L_{\rm SZ} \propto M_{\rm ICM} T_{\rm ICM}$) and potential energy of the hot ICM.  We have found a tight relation $M_{\Delta} \propto L_{\rm SZ}^{3/5}$, and the total cluster mass is directly related to the cluster virial temperature and radius.  Even allowing for the large variation of $f_{\rm ICM}$ in our models, Figure~\ref{fig:potential} shows that the projected half-luminosity radius of the SZE profile, scaled by $L_{\rm SZ}^{1/5} \sim r_{\Delta}$, is well correlated with the potential energy per particle scaled by the cluster virial temperature.  Thus by measuring both $L_{\rm SZ}$ and $r_{\rm SZ}$, one can estimate the level of energy injection now manifest in potential energy of the gas.\\
\clearpage
\clearpage
%
%
\section{Implications of Models}
\label{conc}
In this paper, we have studied the SZE properties of clusters.  Motivated by observations, models reproducing the observed X-ray scaling relations, and the assumptions of spherical symmetry and hydrostatic equilibrium, we have spanned a range of plausible ICM density and temperature profiles.  We expect that the range of parameters we have explored in our models encompasses the properties of real clusters.  While many researchers attempting to understand deviations from self-similar behavior in X-ray properties of clusters have considered a wide range of possible modifications to the gas entropy profiles, we have demonstrated that the SZE luminosity depends only very weakly on the shape and normalization of the entropy profile.  While cooling can significantly reduce the $L_{\rm SZ}$ signal as a large gas fraction is cooled, the average temperature of the remaining gas is surprisingly robust.  Thus, $L_{\rm SZ}$ is an excellent measure of the parameter combination $f_{\rm ICM} M_{\Delta}^{5/3}$.  The robustness of $L_{\rm SZ} \propto M_{\rm ICM} T_{\Delta}$ indicates that for the scope of possible entropy profiles and cluster parameters considered here, any significant injected energy must primarily serve to increase the gas's potential energy by expanding the gaseous region.  However, if non-gravitational processes result in unaccounted cluster-mass dependent gas ejection rates or star formation efficiency, our $L_{\rm SZ} - M$ relation will be distorted.  As shown by \citet{vikhlinin/etal:2005} and discussed in Sec.~\ref{results}, the ICM mass fraction appears lower out to $r_{500}$ in cooler clusters.  Therefore the total ICM mass fraction either depends on cluster mass, or the SZE profiles will be more diffuse in lower mass clusters.  Recent observations of cluster mass/near-infrared K-band luminosity relation \citep{lin/mohr/stanford:2004}, estimates of intracluster light \citep{lin/mohr:2004}, and hydrodynamic simulations finding an ICM mass fraction increasing with mass \citep{kravtsov/nagai/vikhlinin:2005} all suggest the former at some level.  We discuss techniques for constraining $f_{\rm ICM}$ in Sec.~\ref{fbdiscussion}.\\
Of all the variations we have considered, $L_{\rm SZ}$ is more sensitive to the dark matter concentration parameter, the accretion pressure, and the ratio of gas to total mass in the cluster than the shape and normalization of the entropy profile.  As shown in Sec.~\ref{results}, observations support our understanding of dark matter concentration parameters and the gas boundary accretion pressure.\\  
In the context of these models, we have shown the most important and uncertain parameter of the thermal history of the gas is the ICM mass fraction, $f_{\rm ICM}$; properties of the entropy profile induce variations in $L_{SZ}$ clearly bounded by $10\%$ while $L_{SZ} \propto f_{\rm ICM}$.  $L_{SZ}$ shows variations of $\approx 8\%$ from the expected intrinsic scatter in $c_{\Delta}$ \citep{dolag/etal:2004}.  Knowledge of $c_{\Delta}$ and its scatter can be acquired by thorough N-body simulations, and the analysis of \citet{vikhlinin/etal:2005} finds agreement between $\Lambda CDM$ concentration parameters and observations of nearby, relaxed X-ray clusters.  Finally, large deviations of the bounding pressure from our expectations could potentially significantly alter $L_{SZ}$.  The comparison to observations discussed in Sec.~\ref{results} indicates at least mild agreement with our expected values.  Thus we expect the scatter induced by effects discussed here to be bounded by $10\%$.  We caution, however, that our assumption of hydrostatic equilibrium may not hold in the outskirts of the cluster \citep{voit/etal:2002, thomas/etal:2002}.  Furthermore, our modeling is far from extensive: we have ignored cluster asphericity, thermal conduction, intracluster magnetic fields, turbulent support, the presence of a relativistic fluid, and important dynamical events such as large mergers and quasar blast waves.  In an analytic gas model, \citet{ostriker/bode/babul:2005} demonstrate that triaxiality and substructure in simulated dark matter halos do induce some scatter in the cluster observables; they find $\sigma_{\ln y_{100}} \approx 0.3$ for the $y_{100}-M_{100}$ relation when $M_{100} > 10^{14} M_{\sun}$.  \citet{motl/etal:2005} find in hydrodynamic simulations including star formation and supernovae feedback that while major mergers may increase $y_{o}$ by up to a factor of 20, they do not drastically increase $y_{500}$ (see also \citet{randall/sarazin/ricker:2002}).  Furthermore, the level of scatter is similar to our findings: $80\%$ of the simulated cluster mass estimates from the $y_{500}-M_{500}$ relation lie within $+15\%$ to $-10\%$ of the true cluster mass.  The scatter level is nearly constant with redshift back to at least $z = 1.5$, supporting our optimistic view that merging has a small effect on $L_{SZ}$.  Thus we are optimistic that $L_{\rm SZ}$ will be a useful indicator of cluster mass, while SZE profiles coupled with measurements in other wavebands can help us constrain the magnitudes and sources of non-gravitational physics important to understanding clusters.\\
%
%
Our models have indicated a number of tight relations between cluster properties and observables:
\begin{itemize}
\item $L_{SZ} = 1.6 \times 10^{-4} \; {\rm Mpc}^{2} \frac{f_{\rm ICM}}{f_{\rm b}} \frac{M_{\Delta}}{10^{15} h^{-1} M_{\sun}}^{5/3}$
\item $L_{X,cut} \propto y_{o}^{1.2}$ with a normalization independent of cluster mass
\item The net energy injected into the bound ICM can be estimated from the observable $r_{\rm SZ} L_{\rm SZ}^{-1/5}$ as in Figure~\ref{fig:potential} and discussed in Sec.~\ref{energetics}.
\end{itemize}
These relations should now be examined in hydrodynamic simulations, where additional physics can be modeled and departures from hydrostatic equilibrium and spherical symmetry can be examined.\\
\section{Constraining $f_{\rm ICM}$}
\label{fbdiscussion}
Within both our models and current observations, we have shown the most important uncertain parameter in the $L_{\rm SZ} - M_{\Delta}$ relation is the ICM mass fraction, $f_{\rm ICM}$.  Can we observationally determine $f_{\rm ICM}$ as a function of both cluster mass and redshift?\\
SZE surveys can potentially break the $f_{\rm ICM}-M_{\Delta}$ degeneracy inherent in $L_{SZ}$ measurements through the detection of the kinetic Sunyaev-Zel'dovich effect (kSZ).  A cluster's peculiar velocity ${\bf v}$ produces a CMB temperature change given by \citep{sunyaev/zeldovich:1972}
\begin{equation}
\frac{\delta T_{\rm kSZ}}{T_{o}} = \sigma_{Thomson} \int dl n_{e}(l) \left(\frac{-{\bf v}\cdot {\bf n}}{c}\right).
\end{equation}
The kSZ temperature change averaged over cluster $i$, $\delta_{i}/T_{o} = -\sigma_{Thomson} N_{e,i} \; v_{{\rm LOS},i}/c$, is proportional to the total number of cluster electrons and the cluster peculiar velocity projected along the line of sight.  \citet{hernandez/verde/jimenez/spergel:prep} discuss an unbiased estimator of $\delta_{i}$:
\begin{equation}
 \delta_i = \delta T_{CMB,i}^{res} + \delta T_{tSZ,i} + \vnn + \delta
 T_{kSZ,i}^{int} + T_{kSZ,i}.
\label{eq:delta_i}
\end{equation}
The variance of $\delta_{i}$ will therefore have contributions from the desired kSZ signal and a noise term from the residual CMB and thermal SZ signals after subtraction, $\delta T_{CMB,i}^{res} + \delta T_{tSZ,i}$, the instrumental noise, $\vnn$, and internal cluster motions, $\delta T_{kSZ,i}^{int}$.
\begin{equation}
\label{deltavar}
\langle \left(\frac{\delta_{i}}{T_{o}}\right)^{2} \rangle = \sigma_{Thomson}^{2} N_{e,i}^{2} \langle \left(\frac{v_{\rm LOS}}{c}\right)^{2} \rangle + \sigma_{\rm res}^{2}.
\end{equation}
With a good understanding of $\sigma_{\rm res}^{2}$ and given the concordance $\Lambda CDM$ prediction for $\langle v_{\rm LOS}^{2} \rangle$, one may deduce the total electron content and thereby $M_{\rm ICM}$ for a sample of clusters in a certain $L_{\rm SZ}$ and redshift range, thus constraining $f_{\rm ICM}(M,z)$.  This technique will be particularly applicable to SZE surveys, where $\delta T_{\rm kSZ}$ and $L_{\rm SZ}$ can be measured within the same projected area to constrain $f_{\rm ICM}$ without extrapolation in radius.  We caution, however, that the kSZ effect has not yet been detected, and so the systematic errors associated with this measurement are uncertain.\\
As we have demonstrated in Sec.~\ref{phenomsec}, two different models for the cooling and feedback energy distribution can reproduce the density and temperature profiles of observed cool clusters deviating from self-similarity.  Though they differ in $f_{\rm ICM}$ by a factor of 2, these models are indistinguishable out to $\approx 0.85 r_{\rm vir}$, well beyond the typical X-ray detectable region inside $\approx 0.5 r_{\rm vir}$.  In the first (fit 1), non-gravitational heating is well distributed and raises the entropy level of all the cluster gas, and thus increases its potential energy (see discussion in Sec.~\ref{energetics}).  In the second (fit 2), half of the ICM is either cooled or ejected from the cluster, while the remaining hot gas maintains the properties of a gravitationally heated ICM.  These two cluster models do have distinct SZE observables: fit 1 has $L_{SZ} = 0.88 L_{SZ,ss}$ and $r_{SZ} = 0.49$, while fit 2 has $L_{SZ} = 0.57 L_{SZ,ss} = 1.13 L_{SZ,ss}/(f_{\rm ICM}/f_{\rm b})$ and $r_{SZ} = 0.32$.    Both models cause $\lesssim 13\%$ deviation of $L_{SZ}/ (f_{\rm ICM}/f_{\rm b})$ from the self-similar value.  We emphasize, however, that relating the observed thermal SZE profile to $f_{\rm ICM}$ would still require an assumption about the behavior of $\rho(r)$ or $T(r)$ in the outer regions of the cluster.  Since X-ray observations are not sensitive to the outer regions of the cluster, alone they can only place a lower limit on the number of baryons in the ICM and cannot distinguish between these two heating models.  $M_{\Delta}$ can be estimated using other means, such as gravitational lensing or assuming the X-ray observable gas is in hydrostatic equilibrium.  These follow-up measurements could also be used to calibrate the $L_{\rm SZ}-M_{\Delta}$ relation and as a consistency check with the kSZ results.  Because systematics will vary with the technique, multi-wavelength cluster observations will enhance our ability to minimize our assumptions and improve our accuracy in the calibration of the $L_{\rm SZ}-M_{\Delta}$ relation, as well as bring new understanding to the physics of cluster gas.\\
\section{Conclusions}
\label{future}
SZE surveys will soon identify thousands of clusters and measure their photometric redshifts, so we will be able to estimate the gas potential and thermal energy with $L_{\rm SZ}$ and $r_{\rm SZ}$, as well as the cluster mass (see Fig.~\ref{fig:SZ1} and Fig.~\ref{fig:potential}).  Detection of the kSZ signal in these SZE surveys will also constrain the ICM mass fraction, $f_{\rm ICM}(M,z)$.  Follow-up observations in X-ray, optical, and radio can provide checks on the survey $L_{\rm SZ}-M$ relation using independent mass determinations, and could further constrain ICM models through AGN activity-gas energy correlations.  Coupling sufficiently high resolution SZ and X-ray observations could over-constrain the gas density and temperature profiles, and thus provide a consistency check of deprojection techniques.  Sensitive SZE profiles should extend farther into the outer regions of the cluster than X-ray observations allow.\\
This paper has focused on SZE observables from $z=0$ clusters, though SZE surveys will find most of their clusters at $z \sim 0.6$.  However, the models studied here support the self-similar prediction $T_{gas} \sim T_{dark}$ even under considerable variations of the total gas energy.  We expect $T_{dark}$ to scale with the overdensity $\left(\Delta(z) \rho_{cr}(z)\right)^{1/3}$, and the dark matter concentration parameter will decrease with redshift as suggested by Eqn.~\ref{conceqn}: by a factor of $\sim 1.5$ at $z=1$.  Appendix A suggests that changes in $L_{\rm SZ}$ induced by a decrease in $c_{\Delta}$ with redshift is limited to less than $15\%$.  The range of entropy normalizations examined in this paper is larger than the expected characteristic entropy decrease by $\sim 20\%$ at $z=1$ under self-similar evolution.  Because we have found a very weak dependence of $L_{\rm SZ}$ on the entropy profile, we are optimistic that the redshift evolution of the $L_{\rm SZ}-M$ relation will be determined by the more easily understood dark matter evolution; this assumption has been verified in numerical simulations including galaxy formation \citep{nagai:prep}.  In this case the $L_{\rm SZ}-M$ evolution is characterized by relatively few parameters, and so can hopefully be internally calibrated in SZE surveys (\citealt{hu:2003, majumdar/mohr:2004, lima/hu:2004, lima/hu:2005}; but see also \citealt{francis/bean/kosowsky:prep}).  However, we caution that baryon physics may induce a redshift dependent ICM mass fraction $f_{\rm ICM}(M, z)$ that produces a deviation from simple self-similar redshift evolution.  The observed ICM mass fraction in clusters appears to depend on cluster mass and radius.  In Sec.~\ref{fbdiscussion} we proposed the use of the kSZ signal to constrain $f_{\rm ICM}(M,z)$.  Understanding the variation in ICM mass fraction will deepen our understanding of cluster physics and enable the use of clusters as cosmological probes.  Finally, we may hope to probe the gas's thermal history by measuring the thermal and potential energies as a function of redshift through SZE observation.\\

\acknowledgments
BAR acknowledges support from the National Science Foundation Graduate Research Fellowship and useful discussions with Mark Birkinshaw, Jeremiah Ostriker, and Paul Bode.  DNS acknowledges support from NASA Astrophysics Theory Program NNG04GK55G and NSF PIRE grant OISE-0530095.
\appendix
\section{Appendix A}
We list SZE and X-ray observables computed from the models considered in the paper. $r_{\rm ac}$ is the accretion shock radius, given in units of the dark matter virial radius, $r_{\Delta}$.  $r_{\rm SZ}$ and $r_{X}$ are the radii projected on the sky enclosing half the SZE/X-ray luminosity, and also given in units of $r_{\Delta}$.  $y_{o}$ is the unitless Compton parameter measured along the line of sight through the center of the cluster.  We report $L_{\rm SZ}$ in terms of $L_{\rm SZ,ss}$, the value expected from the self-similar collapse model:
\begin{equation}
\label{lszss}
L_{\rm SZ,ss} = \frac{\sigma_{Thomson}}{m_e c^{2}} \frac{f_{\rm b} M_{\Delta}}{\mu m_{p}} T_{\Delta} = 1.71 \times 10^{-4} \; {\rm Mpc}^{2} \frac{M_{\Delta}}{10^{15} h^{-1} M_{\sun}}^{5/3},
\end{equation}
where $f_{\rm b}$ is the universal baryon fraction and $T_{\Delta}$ is the characteristic temperature of the dark matter potential.  $L_{X}$ is the total X-ray luminosity (see Eqn.~\ref{xraylum}), and $L_{X, \rm cut}$ is computed by integrating the X-ray luminosity outside a projected radius of $0.05 r_{\Delta}$.  The X-ray luminosities were computed with the cooling function found in \citet{sutherland/dopita:1993}, which includes emission at all wavelengths.  We also compute the average potential (PE), thermal ($3/2 kT$), and total energy (E) per particle in keV according to the approximation made in Section \ref{energetics}.  $1-f_{\rm ICM}/f_{\rm b}$ is the gas fraction that has been cooled or ejected from the cluster.\\
The tables below list first our baseline model without heating and cooling, followed by models including preheating and cooling with various values of $K_{\rm preheat}$ given in units of ${\rm erg \; cm^2 \; g^{-5/3}}$.  The parameters for the  ``self-similar'' phenomenological model for the $10^{15} h^{-1} M_{\sun}$ cluster are $c_{\Delta} = 8.5$, $w_{\rm accr} = 1$ (the ``fudge factor'' on the mass accretion rate), $s_{1} = s_{2} = 1.1$ (entropy exponents), $x_{core} = 0$ (entropy core radius), $K_{\rm max} = K_{100}$, $K_{\rm min} = 0.01 K_{\rm max}$, and $f_{\rm b} = 0.13$.  The ``self-similar'' parameters are the same for the $10^{14} h^{-1} M_{\sun}$ and $10^{13} h^{-1} M_{\sun}$ clusters except $c_{\Delta} = 10.4$ and $c_{\Delta} = 12.6$ as used in the preheat model of \citet{voit/etal:2003}.  Fit 1 in Table~\ref{table15} (see Sec.~\ref{phenomsec} for discussion) has parameters $c_{\Delta} = 7.7$, $w_{\rm accr} = 1$, $s_{1} = s_{2} = 0.7$, $x_{core} = 0$, $K_{\rm max} = 2.5 K_{100}$, $K_{\rm min} = 0.01 K_{\rm max}$, $f_{\rm b} = 0.13$.  Fit 2 has parameters $c_{\Delta} = 7.7$, $w_{\rm accr} = f_{\rm b}/f_{\rm ICM}$, $s_{1} = s_{2} = 0.7$, $x_{core} = 0$, $K_{\rm max} = K_{100} (f_{\rm ICM}/f_{\rm b})^{-2/3}$, $K_{\rm min} = 0.01 K_{\rm max}$, $f_{\rm b} = 0.13$, and $f_{\rm ICM}/f_{\rm b} = 0.5$.\\
%
\clearpage
\begin{deluxetable}{llllllllllll}
\tabletypesize{\scriptsize}
\tablewidth{0pt}
\tablecolumns{12}
\tablecaption{\label{table15} $10^{15} h^{-1} M_{\sun}$ Cluster Models}
\tablehead{
\colhead{Model} & \colhead{$r_{\rm ac}$} & \colhead{$r_{\rm SZ}$} & \colhead{$y_{o}$} & \colhead{$L_{\rm SZ}$} & \colhead{$r_{X}$} & \colhead{$L_{X}$} & \colhead{$L_{X, \rm cut}$} & \colhead{PE} & \colhead{$\frac{3}{2} kT$} & \colhead{E} & \colhead{$1-f_{\rm ICM}/f_{\rm b}$} \\
\colhead{} & \colhead{$(r_{\Delta})$} & \colhead{$(r_{\Delta})$} & \colhead{$(\times 10^{4})$} & \colhead{$(L_{SZ,ss})$} & \colhead{$(r_{\Delta})$} & \multicolumn{2}{c}{$(10^{44} erg \, s^{-1})$} & \multicolumn{3}{c}{(keV)} & \colhead{}}
\startdata
no heat/cool & 0.893 & 0.255 & 3.6 & 0.963 & 0.0293 & 55. & 18.1 & -29.5 & 9.33 & -20.2 & 0 \\
$K_{\rm preheat} = 10^{34}$ & 1. & 0.297 & 1.28 & 0.97 & 0.123 & 11.5 & 9.84 & -25.5 & 9.4 & -16.1 & 0 \\
$K_{\rm ph} = 5 \times 10^{33}$ & 0.946 & 0.272 & 1.94 & 0.965 & 0.0769 & 21.1 & 14.8 & -27.5 & 9.36 & -18.1 & 0\\
$K_{\rm ph} = 4 \times 10^{33}$ & 0.935 & 0.268 & 2.23 & 0.963 & 0.0635 & 26. & 16. & -28. & 9.34 & -18.7 & 0 \\
$K_{\rm ph} = 3.5 \times 10^{33}$ &0.93 & 0.265 & 2.45 & 0.961 & 0.0549 & 29.9 & 16.3 & -28.3 & 9.32 & -19. & 0 \\
$K_{\rm ph} = 3 \times 10^{33}$ & 0.924 & 0.263 & 2.81 & 0.957 & 0.0428 & 36.8 & 16.1 & -28.6 & 9.29 & -19.3 & 0 \\
$K_{\rm ph} = 2 \times 10^{33}$ & 0.913 & 0.263 & 2.65 & 0.939 & 0.0476 & 31.4 & 15.1 & -28.5 & 9.34 & -19.1 & 0.0252 \\
$K_{\rm ph} = 10^{33}$ & 0.902 & 0.261 & 2.66 & 0.928 & 0.0468 & 31.6 & 15.1 & -28.6 & 9.36 & -19.3 & 0.0378 \\
$K_{\rm ph} = 5 \times 10^{32}$ & 0.896 & 0.259 & 2.69 & 0.923 & 0.046 & 32. & 15. & -28.7 & 9.37 & -19.4 & 0.0442 \\
$K_{\rm ph} = 10^{32}$ & 0.892 & 0.258 & 2.71 & 0.919 & 0.0452 & 32.4 & 15. & -28.8 & 9.37 & -19.5 & 0.049 \\
$K_{\rm ph} = 10^{31}$ & 0.891 & 0.258 & 2.67 & 0.918 & 0.0464 & 31.8 & 15. & -28.8 & 9.37 & -19.4 & 0.0501\\
$K_{\rm ph} = 10^{30}$ & 0.891 & 0.258 & 2.67 & 0.918 & 0.0465 & 31.7 & 15. & -28.8 & 9.37 & -19.4 & 0.0502 \\
self similar & 0.806 & 0.227 & 3.41 & 0.97 & 0.0437 & 51.4 & 23.2 & -31.2 & 9.46 & -21.7 & 0 \\
$c_{\Delta} = 3$ & 0.887 & 0.312 & 0.87 & 0.762 & 0.092 & 14.5 & 10.2 & -24.1 & 7.32 & -16.8 & 0 \\
$c_{\Delta} = 7$ & 0.819 & 0.243 & 2.54 & 0.923 & 0.0509 & 38.6 & 19.5 & -29.3 & 8.89 & -20.4 & 0 \\
$c_{\Delta} = 11$ & 0.781 & 0.205 & 5.08 & 1.04 & 0.0356 & 74.6 & 28.3 & -33.4 & 10.1 & -23.2 & 0 \\
$w_{\rm accr} = 0.28$ & 0.906 & 0.192 & 3.82 & 0.855 & 0.0406 & 64.9 & 27.5 & -32. & 8.32 & -23.6 & 0 \\
$w_{\rm accr} = 3.5$ & 0.65 & 0.268 & 3.03 & 1.39 & 0.0583 & 40.2 & 22. & -30.8 & 13.3 & -17.5 & 0 \\
$s_{1} = s_{2} = 0.7$ & 0.944 & 0.295 & 1.5 & 0.968 & 0.105 & 12.5 & 9.41 & -25.8 & 9.42 & -16.4 & 0 \\
$s_{1} = s_{2} = 1.5$ & 0.669 & 0.167 & 6.42 & 0.946 & 0.0307 & 195. & 53.3 & -37. & 9.14 & -27.9 & 0 \\
$s_{1} = 0$, $x_{c} = 0.1$ & 0.813 & 0.231 & 2.17 & 0.982 & 0.0857 & 28.3 & 21.3 & -29.7 & 9.45 & -20.3 & 0 \\
$K_{\rm min} = 3.5 \times 10^{33}$ & 0.813 & 0.23 & 2.34 & 0.979 & 0.0783 & 31.4 & 22.5 & -30.2 & 9.5 & -20.7 & 0 \\
$K_{\rm max} = K_{200}$ & 0.719 & 0.204 & 3.93 & 0.97 & 0.0419 & 67.7 & 29.6 & -33. & 9.48 & -23.5 & 0 \\
$K_{\rm max} = 1.5 K_{100}$ & 0.988 & 0.274 & 2.64 & 0.964 & 0.047 & 31.2 & 14.9 & -27.8 & 9.31 & -18.5 & 0 \\
$K_{\rm max} = 2 K_{100}$ & 1.15 & 0.316 & 2.21 & 0.951 & 0.0493 & 22.3 & 11. & -25.7 & 9.24 & -16.5 & 0 \\
$K_{\rm max} = 2.5 K_{100}$ & 1.3 & 0.353 & 1.93 & 0.936 & 0.0511 & 17.2 & 8.73 & -24.2 & 9.16 & -15. & 0 \\
$K_{\rm max} = 3 K_{100}$ & 1.43 & 0.385 & 1.71 & 0.926 & 0.0526 & 13.5 & 7. & -22.5 & 8.92 & -13.6 & 0 \\
Cooling threshold & 0.993 & 0.265 & 1.78 & 0.871 & 0.0829 & 18.2 & 13.2 & -28.3 & 9.68 & -18.6 & 0.128 \\
Sec.~\ref{phenomsec} fit 1 & 1.56 & 0.488 & 0.65 & 0.882  & 0.144 & 3.01 & 2.45 & -18.3 & 8.62 & -9.67 & 0 \\
Sec.~\ref{phenomsec} fit 2 & 0.838 & 0.32 & 0.66 & 0.565 & 0.130 & 2.91 & 2.33 & -25.6 & 10.9 & -14.7 & 0.5 \\
$\gamma  = 1.2$, $\rho_{2} = 3.96 \rho_{1}$ & 0.875 & 0.25 & 1.92 & 0.981 & 0.0927 & 21.5 & 16.1 & -28.5 & 9.51 & -19. & 0 \\
$\gamma  = 1.2$, $\rho_{2} = 2.53 \rho_{1}$ & 1.2 & 0.341 & 1.23 & 0.953 & 0.108 & 9.25 & 7.31 & -23.7 & 9.24 & -14.4 & 0 \\
$\gamma  = 1.2$, $\rho_{2} = 1.74 \rho_{1}$ & 1.5 & 0.421 & 0.914 & 0.917 & 0.117 & 5.26 & 4.26 & -20.5 & 8.89 & -11.6 & 0 \\
$\gamma  = 1.2$, $\rho_{2} = 1.05 \rho_{1}$ & 1.83 & 0.483 & 0.758 & 0.866 & 0.119 & 3.64 & 2.96 & -18.3 & 8.4 & -9.86 & 0 
\enddata
\end{deluxetable}
\begin{deluxetable}{llllllllllll}
\tabletypesize{\scriptsize}
\tablewidth{0pt}
\tablecolumns{12}
\tablecaption{\label{table14} $10^{14} h^{-1} M_{\sun}$ Cluster Models}
\tablehead{
\colhead{Model} & \colhead{$r_{\rm ac}$} & \colhead{$r_{\rm SZ}$} & \colhead{$y_{o}$} & \colhead{$L_{\rm SZ}$} & \colhead{$r_{X}$} & \colhead{$L_{X}$} & \colhead{$L_{X, \rm cut}$} & \colhead{PE} & \colhead{$\frac{3}{2} kT$} & \colhead{E} & \colhead{$1-f_{\rm ICM}/f_{\rm b}$} \\
\colhead{} & \colhead{$(r_{\Delta})$} & \colhead{$(r_{\Delta})$} & \colhead{$(\times 10^{5})$} & \colhead{$(L_{SZ,ss})$} & \colhead{$(r_{\Delta})$} & \multicolumn{2}{c}{$(10^{43} erg \, s^{-1})$} & \multicolumn{3}{c}{(keV)} & \colhead{}}
\startdata
no heat/cool & 1.03 & 0.227 & 5.18 & 0.932 & 0.0227 & 54.6 & 13.2 & -6.63 & 1.94 & -4.68 & 0 \\
$K_{\rm preheat} = 10^{34}$ & 1.47 & 0.436 & 0.446 & 0.846 & 0.326 & 1.29 & 1.24 & -3.9 & 1.77 & -2.13 & 0 \\
$K_{\rm ph} = 5 \times 10^{33}$ & 1.26 & 0.325 & 0.888 & 0.91 & 0.197 & 3.48 & 3.23 & -4.87 & 1.9 & -2.97 & 0 \\
$K_{\rm ph} = 4 \times 10^{33}$ & 1.22 & 0.3 & 1.1 & 0.92 & 0.165 & 4.72 & 4.26 & -5.15 & 1.92 & -3.23 & 0 \\
$K_{\rm ph} = 3.5 \times 10^{33}$ & 1.19 & 0.288 & 1.24 & 0.924 & 0.148 & 5.66 & 5.02 & -5.31 & 1.93 & -3.38 & 0 \\
$K_{\rm ph} = 3 \times 10^{33}$ & 1.17 & 0.276 & 1.45 & 0.929 & 0.129 & 7.04 & 6.07 & -5.5 & 1.94 & -3.55 & 0 \\
$K_{\rm ph} = 2 \times 10^{33}$ & 1.11 & 0.251 & 2.28 & 0.932 & 0.0821 & 13.8 & 10.1 & -5.98 & 1.95 & -4.03 & 0 \\
$K_{\rm ph} = 10^{33}$ & 1.05 & 0.249 & 2.95 & 0.852 & 0.0439 & 17.5 & 7.95 & -6.13 & 1.96 & -4.18 & 0.0904 \\
$K_{\rm ph} = 5 \times 10^{32}$ & 1.03 & 0.244 & 3.05 & 0.828 & 0.0408 & 18.4 & 7.9 & -6.23 & 1.97 & -4.26 & 0.12 \\
$K_{\rm ph} = 10^{32}$ & 1. & 0.241 & 2.95 & 0.808 & 0.0425 & 17.4 & 7.7 & -6.28 & 1.97 & -4.3 & 0.144 \\
$K_{\rm ph} = 10^{31}$ & 1. & 0.24 & 2.93 & 0.804 & 0.0429 & 17.2 & 7.66 & -6.29 & 1.97 & -4.31 & 0.149 \\
$K_{\rm ph} = 10^{30}$ & 1. & 0.24 & 2.99 & 0.804 & 0.041 & 17.8 & 7.67 & -6.3 & 1.97 & -4.33 & 0.15 \\
self similar & 0.838 & 0.192 & 4.91 & 0.957 & 0.0357 & 48.4 & 18.4 & -7.18 & 2. & -5.17 & 0 \\
$c_{\Delta} = 4.5$ & 0.919 & 0.261 & 1.47 & 0.753 & 0.0617 & 18.4 & 10.5 & -5.75 & 1.58 & -4.17 & 0 \\
$c_{\Delta} = 8.75$ & 0.85 & 0.205 & 3.74 & 0.909 & 0.04 & 38. & 16. & -6.76 & 1.88 & -4.88 & 0 \\
$c_{\Delta} = 13$ & 0.813 & 0.173 & 7.07 & 1.03 & 0.0303 & 66.3 & 21.4 & -7.65 & 2.14 & -5.5 & 0 \\
$w_{\rm accr} = 0.28$ & 0.912 & 0.168 & 5.39 & 0.896 & 0.0346 & 58.9 & 21.5 & -7.35 & 1.86 & -5.49 & 0 \\
$w_{\rm accr} = 3.5$ & 0.713 & 0.236 & 4.26 & 1.19 & 0.0401 & 35.5 & 15.2 & -6.97 & 2.46 & -4.5 & 0 \\
$s_{1} = s_{2} = 0.7$ & 1. & 0.263 & 1.98 & 0.936 & 0.0929 & 9.41 & 6.67 & -5.75 & 1.94 & -3.8 & 0 \\
$s_{1} = s_{2} = 1.5$ & 0.681 & 0.133 & 9.4 & 0.947 & 0.0287 & 200. & 45.7 & -8.75 & 1.98 & -6.77 & 0 \\
$s_{1} = 0$, $x_{c} = 0.1$ & 0.85 & 0.198 & 2.91 & 0.968 & 0.0809 & 23. & 16.8 & -6.83 & 2.02 & -4.81 & 0 \\
$K_{\rm min} = 3.5 \times 10^{33}$ & 0.95 & 0.261 & 1.21 & 0.948 & 0.186 & 6.35 & 5.85 & -5.67 & 1.99 & -3.69 & 0 \\
$K_{\rm max} = K_{200}$ & 0.744 & 0.173 & 5.65 & 0.964 & 0.034 & 63.9 & 23.3 & -7.62 & 2.03 & -5.59 & 0 \\
$K_{\rm max} = 1.5 K_{100}$ & 1.04 & 0.229 & 3.83 & 0.936 & 0.0387 & 29.8 & 12.2 & -6.43 & 1.95 & -4.48 & 0 \\
$K_{\rm max} = 2 K_{100}$ & 1.21 & 0.26 & 3.21 & 0.916 & 0.0409 & 21.1 & 9.02 & -5.91 & 1.9 & -4.01 & 0 \\
$K_{\rm max} = 2.5 K_{100}$ & 1.38 & 0.287 & 2.81 & 0.896 & 0.0428 & 16.3 & 7.21 & -5.54 & 1.86 & -3.68 & 0 \\
$K_{\rm max} = 3 K_{100}$ & 1.54 & 0.31 & 2.54 & 0.874 & 0.0444 & 13.5 & 6.14 & -5.29 & 1.84 & -3.45 & 0 \\
$w_{\rm accr} =  1.78$ & 0.787 & 0.21 & 4.62 & 1.03 & 0.0369 & 42.5 & 16.8 & -7.08 & 2.15 & -4.93 & 0 \\
$K_{\rm max} \sim T_{\Delta}^{2/3}$ & 1.1 & 0.24 & 3.6 & 0.929 & 0.0395 & 26.3 & 10.9 & -6.25 & 1.94 & -4.32 & 0 \\
$K_{\rm max} \sim T_{\Delta}^{2/3}$, & 0.838 & 0.192 & 2.28 & 0.444 & 0.0357 & 10.4 & 3.97 & -7.18 & 2. & -5.17 & 0.536 \\
\phm{000} $f_b \sim T_{\Delta}^{1/2}$ & & & & & & & & & & & \\
Cooling threshold & 0.97 & 0.251 & 1.34 & 0.705 & 0.111 & 5.42 & 4.43 & -6. & 2.05 & -3.95 & 0.282 \\
$\gamma  = 1.2$, $\rho_{2} = 3.87 \rho_{1}$ & 0.975 & 0.214 & 2.87 & 0.958 & 0.0751 & 20.1 & 13.5 & -6.56 & 2. & -4.56 & 0 \\
$\gamma  = 1.2$, $\rho_{2} = 2.64 \rho_{1}$ & 1.25 & 0.263 & 2.13 & 0.923 & 0.0821 & 11.3 & 7.97 & -5.75 & 1.93 & -3.82 & 0 \\
$\gamma  = 1.2$, $\rho_{2} = 1.74 \rho_{1}$ & 1.55 & 0.309 & 1.7 & 0.883 & 0.0866 & 7.3 & 5.26 & -5.12 & 1.84 & -3.27 & 0 \\
$\gamma  = 1.2$, $\rho_{2} = 1.03 \rho_{1}$ & 1.83 & 0.332 & 1.53 & 0.845 & 0.0871 & 5.95 & 4.3 & -4.78 & 1.77 & -3.01 & 0
\enddata
\end{deluxetable}
\begin{deluxetable}{llllllllllll}
\tabletypesize{\scriptsize}
\tablewidth{0pt}
\tablecolumns{12}
\tablecaption{\label{table13} $10^{13} h^{-1} M_{\sun}$ Cluster Models}
\tablehead{
\colhead{Model} & \colhead{$r_{\rm ac}$} & \colhead{$r_{\rm SZ}$} & \colhead{$y_{o}$} & \colhead{$L_{\rm SZ}$} & \colhead{$r_{X}$} & \colhead{$L_{X}$} & \colhead{$L_{X, \rm cut}$} & \colhead{PE} & \colhead{$\frac{3}{2} kT$} & \colhead{E} & \colhead{$1-f_{\rm ICM}/f_{\rm b}$} \\
\colhead{} & \colhead{$(r_{\Delta})$} & \colhead{$(r_{\Delta})$} & \colhead{$(\times 10^{6})$} & \colhead{$(L_{SZ,ss})$} & \colhead{$(r_{\Delta})$} & \multicolumn{2}{c}{$(10^{42} erg \, s^{-1})$} & \multicolumn{3}{c}{(keV)} & \colhead{}}
\startdata
no heat/cool & 1.15 & 0.196 & 7.58 & 0.942 & 0.0194 & 94.4 & 17.8 & -1.52 & 0.424 & -1.1 & 0 \\
$K_{\rm preheat} = 5 \times 10^{33}$ & 2.43 & 0.666 & 0.2 & 0.663 & 0.743 & 0.467 & 0.461 & -0.539 & 0.298 & -0.241 & 0 \\
$K_{\rm ph} = 4 \times 10^{33}$ & 2.06 & 0.574 & 0.273 & 0.717 & 0.622 & 0.717 & 0.706 & -0.632 & 0.323 & -0.309 & 0 \\
$K_{\rm ph} = 3.5 \times 10^{33}$ & 1.99 & 0.517 & 0.324 & 0.717 & 0.554 & 0.901 & 0.885 & -0.679 & 0.323 & -0.357 & 0 \\
$K_{\rm ph} = 3 \times 10^{33}$ & 1.79 & 0.464 & 0.404 & 0.75 & 0.483 & 1.22 & 1.2 & -0.751 & 0.337 & -0.413 & 0 \\
$K_{\rm ph} = 2 \times 10^{33}$ & 1.43 & 0.347 & 0.764 & 0.853 & 0.312 & 2.91 & 2.78 & -0.967 & 0.384 & -0.584 & 0 \\
$K_{\rm ph} = 10^{33}$ & 1.21 & 0.214 & 4.85 & 0.769 & 0.0318 & 35.5 & 11.9 & -1.44 & 0.422 & -1.02 & 0.18 \\
$K_{\rm ph} = 5 \times 10^{32}$ & 1.12 & 0.238 & 2.69 & 0.642 & 0.0387 & 10.6 & 4.51 & -1.34 & 0.429 & -0.907 & 0.326 \\
$K_{\rm ph} = 10^{32}$ & 1.02 & 0.223 & 2.53 & 0.579 & 0.0424 & 9.51 & 4.25 & -1.4 & 0.437 & -0.96 & 0.403 \\
$K_{\rm ph} = 10^{31}$ & 1. & 0.22 & 2.58 & 0.566 & 0.0394 & 9.82 & 4.17 & -1.41 & 0.439 & -0.974 & 0.419 \\
$K_{\rm ph} = 10^{30}$ & 1. & 0.22 & 2.55 & 0.565 & 0.0404 & 9.62 & 4.16 & -1.41 & 0.439 & -0.974 & 0.42 \\
self similar & 0.856 & 0.163 & 7.13 & 0.985 & 0.0303 & 77.6 & 25.6 & -1.67 & 0.444 & -1.22 & 0 \\
$c_{\Delta} = 7$ & 0.919 & 0.21 & 2.82 & 0.809 & 0.0438 & 42.3 & 19.3 & -1.4 & 0.364 & -1.03 & 0 \\
$c_{\Delta} = 11.5$ & 0.863 & 0.17 & 6.07 & 0.954 & 0.0322 & 68.9 & 24.2 & -1.6 & 0.425 & -1.18 & 0 \\
$c_{\Delta} = 16$ & 0.831 & 0.146 & 10.6 & 1.07 & 0.026 & 102. & 28.6 & -1.8 & 0.484 & -1.32 & 0 \\
$w_{\rm accr} = 0.28$ & 0.912 & 0.148 & 7.68 & 0.956 & 0.0301 & 90.9 & 29.4 & -1.71 & 0.427 & -1.28 & 0 \\
$w_{\rm accr} = 3.5$ & 0.762 & 0.197 & 6.31 & 1.1 & 0.0313 & 59.1 & 20.6 & -1.62 & 0.501 & -1.12 & 0 \\
$s_{1} = s_{2} = 0.7$ & 1.05 & 0.236 & 2.69 & 0.941 & 0.0906 & 12.9 & 8.9 & -1.31 & 0.425 & -0.885 & 0 \\
$s_{1} = s_{2} = 1.5$ & 0.681 & 0.109 & 13.7 & 0.993 & 0.0276 & 334. & 66.8 & -2.07 & 0.446 & -1.62 & 0 \\
$s_{1} = 0$, $x_{c} = 0.1$ & 0.875 & 0.172 & 3.91 & 0.995 & 0.0809 & 32.1 & 23.3 & -1.58 & 0.449 & -1.13 & 0 \\
$K_{\rm max} = K_{200}$ & 0.756 & 0.149 & 8.18 & 0.998 & 0.0287 & 102. & 32.1 & -1.77 & 0.45 & -1.32 & 0 \\
$K_{\rm max} = 1.5 K_{100}$ & 1.08 & 0.193 & 5.63 & 0.954 & 0.0335 & 48.4 & 17.5 & -1.51 & 0.432 & -1.08 & 0 \\
$K_{\rm max} = 2 K_{100}$ & 1.26 & 0.217 & 4.72 & 0.927 & 0.0361 & 34.2 & 13.1 & -1.39 & 0.417 & -0.971 & 0 \\
$K_{\rm max} = 2.5 K_{100}$ & 1.44 & 0.237 & 4.13 & 0.902 & 0.0383 & 26.3 & 10.6 & -1.3 & 0.406 & -0.895 & 0 \\
$K_{\rm max} = 3 K_{100}$ & 1.6 & 0.253 & 3.69 & 0.881 & 0.0402 & 21.2 & 8.87 & -1.23 & 0.393 & -0.833 & 0 \\
$w_{\rm accr} =  3.16$ & 0.769 & 0.193 & 6.35 & 1.09 & 0.0311 & 60. & 20.7 & -1.61 & 0.489 & -1.12 & 0\\
$K_{\rm max} \sim T_{\Delta}^{2/3}$ & 1.54 & 0.247 & 3.89 & 0.889 & 0.0394 & 23.5 & 9.68 & -1.27 & 0.402 & -0.865 & 0 \\
$K_{\rm max} \sim T_{\Delta}^{2/3}$, & 0.856 & 0.163 & 1.54 & 0.212 & 0.0303 & 3.6 & 1.19 & -1.67 & 0.444 & -1.22 & 0.785 \\
\phm{000} $f_b \sim T_{\Delta}^{1/2}$ & & & & & & & & & & & \\
$\gamma  = 1.2$, $\rho_{2} = 3.87 \rho_{1}$ & 1.05 & 0.177 & 4.49 & 0.981 & 0.0643 & 35. & 21.3 & -1.55 & 0.442 & -1.11 & 0\\
$\gamma  = 1.2$, $\rho_{2} = 2.72 \rho_{1}$ & 1.3 & 0.205 & 3.61 & 0.946 & 0.0684 & 22.9 & 14.4 & -1.4 & 0.426 & -0.973 & 0 \\
$\gamma  = 1.2$, $\rho_{2} = 1.80 \rho_{1}$ & 1.58 & 0.23 & 3.06 & 0.909 & 0.0711 & 16.6 & 10.7 & -1.28 & 0.409 & -0.873 & 0 \\
$\gamma  = 1.2$, $\rho_{2} = 1.06 \rho_{1}$ & 1.83 & 0.24 & 2.85 & 0.881 & 0.0715 & 14.4 & 9.32 & -1.22 & 0.397 & -0.827 & 0
\enddata
\end{deluxetable}
\clearpage
\bibliographystyle{apj}
\bibliography{apj-jour,wmapbeth}
\end{document}